\documentclass[aps,pre,reprint,groupedaddress,amsmath,amssymb,floatfix]{revtex4-2}
\usepackage{graphicx}
\usepackage{amsthm}
\usepackage{microtype}
\usepackage[hidelinks]{hyperref}
\newtheorem{proposition}{Proposition}
\DeclareMathOperator{\sech}{sech}
\DeclareMathOperator{\erfc}{erfc}
\newcommand{\umax}{u_{\max}}
\newcommand{\dd}{\mathrm{d}}

\begin{document}

\title{Exact solution of the two-dimensional one-component plasma on the truncated Beltrami pseudosphere}

\author{Josep Batle}
\email{batlequantum@gmail.com}
\affiliation{CRISP -- Centre de Recerca Independent de sa Pobla, 07420 sa Pobla, Balearic Islands, Spain}

\date[]{}

\begin{abstract}
The two-dimensional one-component plasma, $N$ point charges in a neutralizing background interacting through the logarithmic Coulomb potential, is solved exactly at the coupling $\Gamma=2$ on the truncated Beltrami pseudosphere, the surface of constant negative curvature $-1/a^{2}$ generated by the tractrix, bounded by its rim and by a second parallel at the value $\umax$ of the parameter $u$ that runs along the meridian from the rim. A conformal change of coordinates maps the surface onto a flat cylinder, on which the particles interact through the flat-cylinder potential and the curvature enters only through the area element; the partition function becomes a product of incomplete gamma functions, the density is explicit and all correlations are determinantal. The thermodynamic limit is taken at fixed background density on surfaces of the same shape and growing size, $\umax$ being fixed and $a$ growing as $\sqrt N$, as is done on the sphere; since the logarithmic interaction has no intrinsic length, this limit is equivalent to letting $N$ grow on a surface of fixed size. The free energy is obtained through the constant term with every coefficient in closed form: a bulk term proportional to $N$, the flat hard-wall surface tension times the total length of the two boundaries, and a shape-dependent constant, with no term in $\ln N$. This is the prediction of the conformal anomaly for a surface of Euler characteristic zero, tested here on a surface on which the integrated Gaussian curvature and the integrated boundary curvature are separately nonzero and cancel; in the conformal coordinates the model is a cylinder plasma with a specific axial potential, and the expansion coincides term by term with the general cylinder result recently obtained by Shen and Forrester, whose universal constants are the quadratures of the two boundary layers. At both walls the density tends to the flat contact value $n_{b}\ln2$, from opposite sides, although their circumferences differ by the factor $\cosh\umax$. The asymptotic results are compared with the exact solution up to $N=4096$, and Monte Carlo simulations at $\Gamma=1$, $4$ and $6$, validated against the exact density at $\Gamma=2$, show how the two boundary layers evolve with the coupling.
\end{abstract}

\maketitle

\begin{figure*}[t]
\centering
\includegraphics[width=\textwidth]{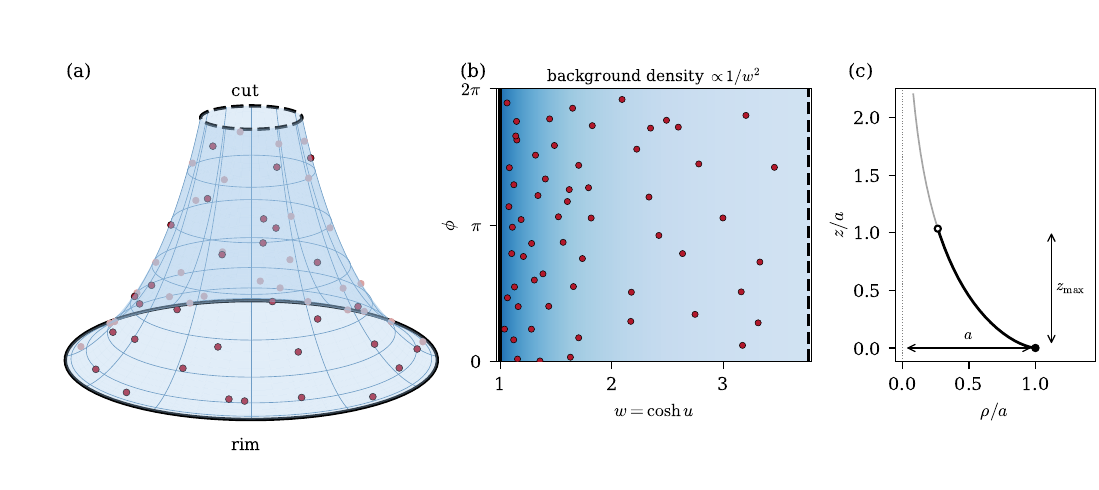}
\caption{The system. (a) The truncated Beltrami pseudosphere for $\umax=2$, with the rim $u=0$ (solid) and the cut $u=\umax$ (dashed), carrying $N=60$ point charges in an equilibrium configuration at $\Gamma=2$ obtained by Metropolis sampling (dark: front face; light: back face). The surface also carries a uniform neutralizing background (blue tint). (b) The same configuration in the conformal coordinates $(w,\phi)$ of Sec.~\ref{sec:surface}, in which the surface is the flat cylinder $1\le w\le W=\cosh\umax$, $0\le\phi<2\pi$; the background, uniform on the curved surface, is concentrated near the rim with density proportional to $1/w^{2}$ (shading), and the charges follow it. (c) The tractrix profile $z(\rho)$ of the surface, Eq.~(\ref{eq:embed}); the thick part is the truncated surface between the rim ($\bullet$) and the cut ($\circ$), and $z_{\max}=a(\umax-\tanh\umax)$ is the height of the cut above the rim in the embedding.}
\label{fig:system}
\end{figure*}

\section{Introduction}\label{sec:intro}

Charged particles confined to a surface are common in physics: electrons on the surface of liquid helium, ions adsorbed on a membrane or a colloid, charges trapped at an interface. When the electric field lines are also confined to the surface, the potential created by a point charge is not $1/r$ but the solution of Poisson's equation in two dimensions, the logarithm $-\ln r$ of the distance. A useful way to picture this is to think of each particle as an infinitely long, uniformly charged line perpendicular to the plane: the field of such a line decays as $1/r$ and its potential is logarithmic. Systems of charges interacting in this way are called two-dimensional Coulomb systems \cite{Jancovici1995,Forrester1998,ForresterBook}. The logarithmic interaction is long ranged, so that these systems screen, carry surface charges and obey sum rules exactly as their three-dimensional counterparts do, while being simple enough to be solved exactly in some cases.

The simplest of them is the one-component plasma, or jellium: $N$ identical point charges $q$ moving in a uniform background of the opposite charge, which makes the whole neutral and holds the charges together. In equilibrium at temperature $T$ the model depends on a single dimensionless parameter, the coupling $\Gamma=q^{2}/k_{B}T$. In two dimensions $q^{2}$ is an energy, and the coupling compares the electrostatic energy with the thermal energy. At small $\Gamma$ the plasma is a weakly correlated gas; at large $\Gamma$ the repulsion wins and the charges crystallize, near $\Gamma\simeq140$ in the plane \cite{CLWH1982}. At the particular value $\Gamma=2$, as Jancovici and Alastuey showed in 1981 \cite{Jancovici1981,AlastueyJancovici1981}, the plasma can be solved exactly: the partition function and all the correlation functions are obtained in closed form. The reason, which is explained in detail in Sec.~\ref{sec:whygamma2}, is that at this coupling the Boltzmann factor of the pair interactions is the squared modulus of a determinant, the same mathematical structure that describes noninteracting fermions. The same structure appears in random-matrix theory, where it describes the eigenvalues of the Ginibre ensemble of complex matrices \cite{Ginibre1965,ForresterBook}, and in Laughlin's analogy between the fractional quantum Hall effect and a classical plasma \cite{Laughlin1983}.

The exact solution survives when the plane is replaced by a curved surface, provided the metric of the surface can be written as a flat metric multiplied by a position-dependent factor, because the two-dimensional Coulomb potential keeps its flat form under such a change of metric. The model has been solved in this way on the cylinder \cite{CFS1983}, on the sphere \cite{Caillol1981}, on discs of the hyperbolic plane \cite{JancoviciTellez1998,FJT2003}, and on Flamm's paraboloid, the spatial section of the Schwarzschild geometry \cite{FantoniTellez2008}; Fantoni has reviewed these solutions in a common language \cite{FantoniReview}. They have served two purposes. One is to see how the curvature of the surface changes the structure of a Coulomb fluid. The other is to test a remarkable prediction about finite systems. A two-dimensional Coulomb system of linear size $L$ has a free energy that contains, besides the bulk term proportional to the area and the boundary term proportional to the perimeter, a term proportional to $\ln L$ whose coefficient is universal: it depends neither on the coupling nor on the microscopic details, but only on the Euler characteristic $\chi$ of the surface, the topological number that is $2$ for a sphere, $1$ for a disc and $0$ for a cylinder or an annulus \cite{JMP1994,JancoviciTellez1996}. The prediction comes from regarding the Coulomb system as a critical system, whose free energy contains a logarithmic term fixed by the conformal anomaly \cite{CardyPeschel1988}; it has been verified on the disc, on the sphere and on several flat domains \cite{JMP1994,JancoviciTellez1996,TellezForrester1999,SalazarTellez2016}, and it governs the large-$N$ expansion of random normal matrices on curved surfaces \cite{ZabrodinWiegmann2006,Klevtsov2014,CanLaskinWiegmann2014,FerrariKlevtsov2014}.

The anomaly couples to the geometry through two quantities: the Gaussian curvature integrated over the surface and the curvature of the boundary integrated along it. The Gauss--Bonnet theorem states that their sum is $2\pi\chi$. On every surface on which the prediction has been tested, however, one of the two is zero: flat domains have no Gaussian curvature, and the sphere has no boundary. Such tests cannot distinguish the prediction from a formula that weights the two contributions differently. The sharpest test is a surface on which both are present, of order one, and cancel.

The truncated Beltrami pseudosphere is such a surface. It is the trumpet obtained by rotating the tractrix about its asymptote, Fig.~\ref{fig:system}(a). It has the constant negative Gaussian curvature $-1/a^{2}$, the length $a$ being its radius of curvature, and when it is cut along two parallels its two boundary circles have a curvature whose integral is exactly the negative of the integrated Gaussian curvature: its Euler characteristic is zero, like that of any finite cylinder, while neither term vanishes. The surface has a history in geometry: Beltrami used it in 1868 as the first concrete model of the non-Euclidean plane of Lobachevsky \cite{Beltrami1868}, and Hilbert proved that no complete surface of constant negative curvature can be embedded smoothly in ordinary space \cite{Hilbert1901}, and the pseudosphere, consistently, ends at a sharp rim. It is also a physical object. Graphene sheets shaped as Beltrami pseudospheres have been studied as laboratories for quantum fields in curved space \cite{IorioLambiase2014,Taioli2016,Morresi2020}, and truncated pseudospheres have recently been fabricated as organic microlasers by direct laser writing \cite{Girin2025}. The one-component plasma has been solved on the complete hyperbolic plane, which in that literature is also called the pseudosphere, and on its discs bounded by a single circle, by mapping them onto the Poincar\'e disc \cite{JancoviciTellez1998,FJT2003}. Those domains are simply connected. Beltrami's surface is a different object, a strip of the hyperbolic plane rolled up into a tube; the doubly bounded truncated pseudosphere has a different topology and, to the best of our recollection, has not been treated exactly.

The present contribution solves the plasma at $\Gamma=2$ on the truncated pseudosphere and uses the solution to separate three effects that are usually mixed: bulk thermodynamics, boundary physics and topology. The construction is carried out from the Hamiltonian, each step is given its physical reason, and every asymptotic result is compared with the exact solution at finite $N$. One feature distinguishes this geometry from a generic curved domain: the neutrality condition removes the position of the cut from the one-body weights, so that the truncated plasma feels the same confinement as on the untruncated surface. The thermodynamic limit is taken at fixed background density on surfaces of the same shape and growing size, the radius of curvature $a$ growing as $\sqrt N$ at fixed $\umax$; this is the limit taken on the sphere, whose radius grows at fixed density while its integrated curvature remains $4\pi$. The present contribution is divided as follows. Section~\ref{sec:model} defines the model, derives the configurational partition function from the Hamiltonian, explains the role of the coupling and why $\Gamma=2$ is exactly solvable, and describes the expected structure of the free energy of a finite system. Section~\ref{sec:surface} describes the truncated pseudosphere and the conformal coordinate that maps it onto a flat cylinder. Section~\ref{sec:background} computes the potential of the background and states the neutrality relation. Section~\ref{sec:exact} derives the exact solution, and Sec.~\ref{sec:finiteN} describes the plasma at finite $N$. Section~\ref{sec:largeN} defines the thermodynamic limit, derives the expansion of the free energy and compares it with the exact solution. Section~\ref{sec:anomaly} relates the result to the conformal anomaly, and Sec.~\ref{sec:contact} derives the density at the two walls. Section~\ref{sec:gamma} presents Monte Carlo simulations at other couplings. Finally, some conclusions are drawn in Sec.~\ref{sec:conclusions}.

\section{The model}\label{sec:model}

\subsection{Coulomb potential on a surface}

Consider a two-dimensional surface $S$ with a Riemannian metric $g$, so that the squared length of an infinitesimal displacement is $\dd s^{2}=g_{ab}\,\dd x^{a}\dd x^{b}$ in local coordinates $x^{a}$, and the area element is $\dd A=\sqrt{g}\,\dd^{2}x$ with $g=\det g_{ab}$. The two-dimensional Coulomb potential $G(r,r')$ created at the point $r$ by a unit charge at $r'$ is the solution of Poisson's equation on the surface,
\begin{equation}
\Delta_{g}\,G(r,r')=-2\pi\,\delta_{g}(r,r'),\label{eq:green}
\end{equation}
where $\Delta_{g}$ is the Laplace--Beltrami operator, the Laplacian of the curved surface, and $\delta_{g}$ the delta function normalized with respect to $\dd A$. In local coordinates, with $g^{ab}$ the inverse of $g_{ab}$,
\begin{equation}
\Delta_{g}G=\frac{1}{\sqrt g}\,\partial_{a}\bigl(\sqrt g\,g^{ab}\,\partial_{b}G\bigr),\qquad
\delta_{g}(r,r')=\frac{\delta^{2}(x-x')}{\sqrt g},\label{eq:LB}
\end{equation}
so that $\int\delta_{g}\,\dd A=1$. In the plane the solution is $G=-\ln|r-r'|$, and Eq.~(\ref{eq:green}) is Gauss's law with the field lines confined to the surface. Two charges $q$ interact with the energy $q^{2}G$. Gaussian units are used throughout: Poisson's equation has the form (\ref{eq:green}), with no dielectric constant, so that a unit charge at distance $r$ in the plane creates the potential $-\ln r$ and the square of a charge is an energy; the unit in which $r$ is measured fixes the zero of the potential. Temperatures are measured in energy units through $k_{B}T$.

\subsection{Background, Hamiltonian and neutrality}

A set of like charges alone has no equilibrium: the energy decreases without bound as they move apart. The plasma is held together by a background of charge density $-qn_{b}$ per unit area, uniform on the surface and fixed. The background creates the potential
\begin{equation}
v_{b}(r)=-qn_{b}\int_{S}G(r,r')\,\dd A'\label{eq:vbdef}
\end{equation}
and has the electrostatic self-energy
\begin{equation}
E_{bb}=\frac12\int_{S}(-qn_{b})\,v_{b}(r)\,\dd A.\label{eq:Ebbdef}
\end{equation}
The particles have mass $m$ and canonical momenta $p_{a}$ conjugate to the coordinates $x^{a}$. A free particle moving on the surface has the kinetic energy $\tfrac{1}{2m}g^{ab}p_{a}p_{b}$, where $g^{ab}$ is the inverse of the metric. The Hamiltonian of the system is therefore
\begin{equation}
H=\sum_{i=1}^{N}\frac{g^{ab}(r_{i})\,p_{i,a}\,p_{i,b}}{2m}+U(r_{1},\dots,r_{N}),\label{eq:H}
\end{equation}
with the potential energy
\begin{equation}
U=q^{2}\sum_{i<j}G(r_{i},r_{j})+q\sum_{i=1}^{N}v_{b}(r_{i})+E_{bb}.\label{eq:U}
\end{equation}
The three terms are the repulsion between the particles, the attraction of each particle by the background, and the self-energy of the background, respectively. The particles are confined to a region of the surface by hard walls, which forbid them to cross the boundary and do not act on the electric field: the medium has the same dielectric constant on both sides of each wall, so that there are no image charges. The amount of background is fixed by the neutrality condition
\begin{equation}
N=n_{b}A,\label{eq:neutral0}
\end{equation}
$A$ being the area of the region. Neutrality is a physical requirement: a non-neutral system has an energy that grows faster than $N$ and no bulk to speak of.

\subsection{Partition function and the coupling $\Gamma$}

At temperature $T$, with $\beta=1/k_{B}T$, the canonical partition function $\mathcal{Z}$ is the integral of $e^{-\beta H}$ over the phase space, divided by $N!\,h^{2N}$. The Gaussian integral over the two momenta of a particle gives $(2\pi m/\beta)\sqrt g$, and the factor $\sqrt g$ combines with $\dd^{2}x$ into the area element. With the thermal de Broglie wavelength $\Lambda=h/\sqrt{2\pi mk_{B}T}$,
\begin{equation}
\mathcal{Z}=\Lambda^{-2N}Z_{N},\qquad Z_{N}=\frac{1}{N!}\int_{S^{N}}\prod_{i=1}^{N}\dd A_{i}\;e^{-\beta U}.\label{eq:ZNdef}
\end{equation}
The kinetic energy contributes only the factor $\Lambda^{-2N}$, as in flat space, and the curvature of the surface enters the configurational partition function $Z_{N}$ through the area element $\dd A$. The configurational free energy
\begin{equation}
\beta F=-\ln Z_{N}\label{eq:Fdef}
\end{equation}
is the object of study. Since $Z_{N}$ has the dimension of an area to the power $N$, $\beta F$ contains a term $-N\ln a^{2}$ when lengths are measured with a reference length $a$; in all numerical work lengths are in units of $a$, that is, $a=1$, and this term is zero.

All three terms of Eq.~(\ref{eq:U}) carry the factor $q^{2}$, so that
\begin{equation}
\beta U=\Gamma\,\mathcal{U},\qquad \Gamma:=\beta q^{2}=\frac{q^{2}}{k_{B}T},\label{eq:Gamma}
\end{equation}
where $\mathcal{U}$ depends on the positions, on $n_{b}$ and on the geometry, but not on $q$ or $T$. The coupling $\Gamma$ is the only thermodynamic parameter. In contrast with the three-dimensional plasma, where the coupling involves the mean interparticle distance, no length enters $\Gamma$, because the logarithm has no intrinsic scale.

\subsection{Why the coupling $\Gamma=2$ is exactly solvable}\label{sec:whygamma2}

The difficulty of the partition function (\ref{eq:ZNdef}) is the pair term, which couples every particle to every other. In the plane, with the complex coordinates $z_{i}=x_{i}+iy_{i}$, this term is $\prod_{i<j}|z_{i}-z_{j}|^{\Gamma}$, and for $\Gamma=2$ it is the squared modulus of the Vandermonde determinant,
\begin{equation}
\prod_{i<j}|z_{i}-z_{j}|^{2}=\bigl|\det\bigl(z_{i}^{\,k}\bigr)\bigr|^{2},\label{eq:vandermonde}
\end{equation}
with $i=1,\dots,N$ and $k=0,\dots,N-1$. A squared determinant is integrated by Andr\'eief's identity \cite{Andreief1883,ForresterBook}: for any functions $\psi_{k}$ and any measure $\dd\mu$,
\begin{equation}
\frac{1}{N!}\int\prod_{i}\dd\mu(r_{i})\,\bigl|\det\psi_{k}(r_{i})\bigr|^{2}=\det\Bigl(\int\overline{\psi_{k}}\,\psi_{l}\,\dd\mu\Bigr).\label{eq:andreief}
\end{equation}
When the $\psi_{k}$ are orthogonal for $\dd\mu$ the matrix on the right-hand side is diagonal and the partition function is the product $\prod_{k}h_{k}$ of the norms $h_{k}=\int|\psi_{k}|^{2}\dd\mu$; the one-body density is $\sum_{k}|\psi_{k}|^{2}/h_{k}$ times the weight of the measure, and every correlation function is a determinant built from the kernel $K_{N}(r,r')=\sum_{k}\overline{\psi_{k}(r)}\psi_{k}(r')/h_{k}$. These procedures are well known \cite{Jancovici1981,AlastueyJancovici1981,ForresterBook}, and only the steps that are specific to the present surface are written out in Sec.~\ref{sec:exact}.

This structure has a physical reading. The squared modulus of a determinant of one-particle wave functions is the probability density of $N$ noninteracting fermions occupying those wave functions, and in the plane the plasma at $\Gamma=2$ is the probability distribution of a filled lowest Landau level of a charged particle in a magnetic field, which is Laughlin's plasma analogy at filling factor one \cite{Laughlin1983}. At $\Gamma=4,6,\dots$ the pair term is still a polynomial but no longer a squared determinant; its expansion in symmetric functions gives exact results for small $N$ \cite{SamajPercus1995,TellezForrester1999,Samaj2004,SalazarTellez2016} but no closed form, and at non-even $\Gamma$ only numerical methods are available \cite{CLWH1982}. The strategy of the present contribution is to solve the model exactly at $\Gamma=2$, to use the solution as a controlled laboratory for the geometry, and then to use it as a benchmark for simulations at other couplings.

\subsection{Free energy of a finite system}

In a finite region the density is not uniform: it departs from $n_{b}$ near the walls over a distance of the order of the interparticle spacing $n_{b}^{-1/2}$. Accordingly, the free energy of a large system has the structure
\begin{equation}
\beta F=\beta f\,N+\beta\gamma\,L_{b}+\frac{\chi}{12}\ln N+\dots,\label{eq:Fstructure}
\end{equation}
with a bulk free energy $f$ per particle, a surface tension $\gamma$ times the length $L_{b}$ of the boundary, and the universal logarithmic term described in the introduction, written here for a sequence of systems of fixed shape and density, so that $L_{b}\propto\sqrt N$. The origin of the three terms is as follows. The first two are of local origin: the particles in the interior see the same uniform environment, which gives a contribution proportional to their number, and those within a few spacings of the wall see a modified one, which gives a contribution proportional to the length of the boundary; these are the thermodynamic limits of the free energy per particle and of the surface tension, which exist for the Coulomb system at $\Gamma=2$ because the solution is explicit \cite{Jancovici1981,Jancovici1982}. The third term is not local. Jancovici, Manificat and Pisani \cite{JMP1994} obtained it by treating the Coulomb system as a critical system in the sense of Cardy and Peschel \cite{CardyPeschel1988}, with central charge $c=-1$: the free energy of such a system in a domain of linear size $\ell$ contains the term $-(c\chi/6)\ln\ell$, and with $\ell^{2}\propto N$ at fixed density this is $(\chi/12)\ln N$. The dots in Eq.~(\ref{eq:Fstructure}) stand for a constant and for corrections that vanish as $N\to\infty$. Separating these contributions on a curved surface, and testing the logarithmic one when $\chi=0$, is the purpose of the exact solution.

\section{The truncated pseudosphere}\label{sec:surface}

\subsection{Embedding and metric}

The tractrix is the curve traced by a point dragged by a string of length $a$ whose other end moves along a straight line. Its surface of revolution about that line is the pseudosphere. With $u\ge0$ a parameter along the meridian and $\phi$ the azimuth, the surface is
\begin{equation}
\rho(u)=a\sech u,\qquad z(u)=a\,(u-\tanh u),\label{eq:embed}
\end{equation}
where $\rho$ is the distance to the axis; Fig.~\ref{fig:system}(c) shows the profile. At $u=0$ the radius $\rho=a$ is largest; as $u$ grows the surface narrows into an infinitely long cusp. Differentiating (\ref{eq:embed}) gives the metric
\begin{equation}
\dd s^{2}=a^{2}\bigl(\tanh^{2}u\,\dd u^{2}+\sech^{2}u\,\dd\phi^{2}\bigr),\label{eq:metric}
\end{equation}
whose Gaussian curvature is the constant $K=-1/a^{2}$: the length $a$ of the string is the radius of curvature. The coefficient of $\dd u^{2}$ vanishes at $u=0$, so that the surface is not smooth at its widest circle, where the meridians meet at a sharp edge. The classical pseudosphere thus ends at a singular rim, consistent with Hilbert's theorem that no complete smooth surface of constant negative curvature can be embedded in Euclidean three-space \cite{Hilbert1901}. Intrinsically, the surface (\ref{eq:embed}) is a region of the hyperbolic plane rolled up along a family of horocycles. Let us recall that a horocycle is the curve of the hyperbolic plane obtained as the limit of a circle whose radius grows without bound while it remains tangent to a fixed line at a fixed point; it has the constant geodesic curvature $1/a$. The parallels $u=$ const of the pseudosphere are horocycles \cite{IorioLambiase2014}.

The plasma is confined between the rim and a cut at $u=\umax$. The area of this truncated surface is
\begin{equation}
A=\int_{0}^{\umax}\!\!\dd u\int_{0}^{2\pi}\!\!\dd\phi\,a^{2}\tanh u\sech u=2\pi a^{2}\Bigl(1-\frac1W\Bigr),\label{eq:area}
\end{equation}
with $W:=\cosh\umax$. It is bounded by $2\pi a^{2}$ however far the cut is taken: the infinite cusp has a finite area. The proper length of a meridian between the rim and the cut is
\begin{equation}
\ell=a\int_{0}^{\umax}\tanh u\,\dd u=a\ln\cosh\umax,\label{eq:ell}
\end{equation}
which differs from the height $z_{\max}=a(\umax-\tanh\umax)$ of the cut above the rim, since the meridian climbs obliquely. For a narrow band, $\ell\simeq a\umax^{2}/2$ while $z_{\max}\simeq a\umax^{3}/3$. Only the intrinsic quantities, areas and proper lengths, enter the plasma.

\subsection{The two boundaries and the Gauss--Bonnet theorem}

The two boundary circles have the proper lengths
\begin{equation}
L_{\mathrm{rim}}=2\pi a,\qquad L_{\mathrm{cut}}=\frac{2\pi a}{W},\label{eq:lengths}
\end{equation}
and both are horocycles of geodesic curvature $1/a$ in magnitude. Seen from inside the band the rim bends outward and the cut inward. Locally the two walls are therefore identical, which makes the surface a clean test: any difference between them is a global effect of the geometry.

The Gauss--Bonnet theorem states that for a surface $S$ with boundary $\partial S$,
\begin{equation}
\int_{S}K\,\dd A+\oint_{\partial S}\kappa_{g}\,\dd s=2\pi\chi,\label{eq:GBgeneral}
\end{equation}
where $\kappa_{g}$ is the geodesic curvature of the boundary and $\chi$ is the Euler characteristic, a topological invariant. For the truncated pseudosphere,
\begin{equation}
\int_{S}K\,\dd A=-\frac{A}{a^{2}}=-2\pi\Bigl(1-\frac1W\Bigr),\label{eq:GB}
\end{equation}
and the boundary integral is $L_{\mathrm{rim}}/a-L_{\mathrm{cut}}/a=+2\pi(1-1/W)$, the signs being fixed by the orientation. The sum is zero, so that $\chi=0$, while each integral is of order one and depends on $\umax$.

\subsection{The conformal coordinate}

The ratio of the two coefficients of the metric (\ref{eq:metric}) is $\sinh^{2}u$, and the substitution that absorbs it is
\begin{equation}
w=\cosh u,\qquad \dd w=\sinh u\,\dd u,\qquad 1\le w\le W.\label{eq:w}
\end{equation}
In the coordinates $(w,\phi)$ the metric reads
\begin{equation}
\dd s^{2}=\frac{a^{2}}{w^{2}}\bigl(\dd w^{2}+\dd\phi^{2}\bigr):\label{eq:conformal}
\end{equation}
the flat metric of a cylinder of circumference $2\pi$ multiplied by the conformal factor $a^{2}/w^{2}$. The surface is thus conformally equivalent to the flat cylinder $[1,W]\times S^{1}$, with the rim at $w=1$ and the cut at $w=W$, Fig.~\ref{fig:system}(b). The area element is $\dd A=(a^{2}/w^{2})\,\dd w\,\dd\phi$, and the circle $w=$ const has the proper circumference $2\pi a/w$. In the upper half-plane model of the hyperbolic plane, $w$ is the height $y/a$ and $\phi$ the horizontal coordinate $x/a$, identified modulo $2\pi$. The complex coordinate $\zeta=w+i\phi$ will be used throughout.

Two other flat representations are related to this one. The first is the Poincar\'e disc, onto which the complete hyperbolic plane is mapped, and which was used to solve the plasma on that plane and on its discs \cite{JancoviciTellez1998,FJT2003}. The hyperbolic plane is simply connected. Beltrami's surface is the part $w\ge1$ of the upper half-plane with the points $\phi$ and $\phi+2\pi$ identified, and it is doubly connected; in the Poincar\'e disc it is the interior of a horocycle with a periodic identification along it, and the rotations of the surface about its axis are not rotations of the disc. The second representation is the annulus: the exponential $z=e^{\zeta}$ maps the band $1\le w\le W$ onto the annulus $e\le|z|\le e^{W}$, the rotations of the surface become rotations about the origin, and the metric (\ref{eq:conformal}) becomes $\dd s^{2}=a^{2}|\dd z|^{2}/(|z|\ln|z|)^{2}$. The annulus and the cylinder are equivalent descriptions, related by a change of variable, and the annulus form is the one used in Sec.~\ref{sec:validation}. The cylinder is used elsewhere because the surface is a tube with two ends and because its conformal factor is a power of the coordinate, which makes the integrals over $w$ elementary.

\subsection{The Coulomb potential in conformal coordinates}

A conformal factor $\Omega^{2}$ multiplies the metric, and in two dimensions it divides both the Laplacian and the delta function by the same $\Omega^{2}$: $\Delta_{g}=\Omega^{-2}\Delta_{0}$ and $\delta_{g}=\Omega^{-2}\delta_{0}$, the subscript $0$ denoting the flat cylinder. Equation~(\ref{eq:green}) is therefore the same in both metrics.

Equation~(\ref{eq:green}) determines $G$ only up to a function harmonic in each of its two arguments, and this freedom has physical content: such a function adds image charges or a position-dependent self-energy, and defines a different model. It is fixed by stating where the field lives. The walls are mechanical only: they confine the particles and impose no electrostatic boundary condition, neither Dirichlet nor Neumann, so that the field extends beyond them. The field lives on the complete surface of revolution of the tractrix, whose second sheet, $u<0$ in Eq.~(\ref{eq:embed}), meets the first at the rim. With $\tilde w=\cosh u$ for $u\ge0$ and $\tilde w=2-\cosh u$ for $u<0$, its metric is $a^{2}(\dd\tilde w^{2}+\dd\phi^{2})$ divided by $\tilde w^{2}$ on the first sheet and by $(2-\tilde w)^{2}$ on the second, so that the complete surface is conformal to the infinite flat cylinder $-\infty<\tilde w<\infty$; on the sheet that carries the plasma, $\tilde w=w$. The Coulomb potential is taken to be the potential of that cylinder that depends only on $\zeta-\zeta'$. It is invariant under translations along the cylinder, a neutral set of charges creates no field at either of its ends, and it carries no position-dependent self-energy in the flat coordinates. Other choices, such as the Green function of the hyperbolic plane that vanishes at its ideal boundary and is used for discs \cite{JancoviciTellez1998,FJT2003}, differ from it by terms harmonic in each argument and define different models. On the cylinder of circumference $2\pi$ the potential is
\begin{equation}
G(\zeta,\zeta')=-\ln\Bigl|\,2\sinh\frac{\zeta-\zeta'}{2}\Bigr|.\label{eq:G}
\end{equation}
It is periodic in $\phi-\phi'$, behaves as $-\ln|\zeta-\zeta'|$ at short distance, and as $-|w-w'|/2$ when the separation along the axis is large. In terms of the geodesic distance $d$ between two nearby points of coordinate $w$, for which $|\zeta-\zeta'|=wd/a$ by Eq.~(\ref{eq:conformal}), the short-distance form is $G\simeq-\ln(d/a)-\ln w$: the potential (\ref{eq:G}) measures distances in units of $a$ and contains the position-dependent term $-\ln w$. On the cylinder the potential depends only on $\zeta-\zeta'$, which is a further reason for using this representation; in the annulus variable $z=e^{\zeta}$ it reads $G=-\ln|z-z'|+\tfrac12\ln|z|+\tfrac12\ln|z'|$. The planar potential $-\ln|z-z'|$ on the annulus, for which the field lives on the punctured plane, is another of the choices mentioned above: for a neutral system it changes the potential energy (\ref{eq:U}) by the one-body term $(q^{2}/2)\sum_{i}w_{i}$, the contributions of the pairs, of the background and of its self-energy cancelling up to this remainder, and that model is solved by the same method. In real variables,
\begin{equation}
\Bigl|2\sinh\frac{\zeta-\zeta'}{2}\Bigr|^{2}=2\bigl[\cosh(w-w')-\cos(\phi-\phi')\bigr].\label{eq:Greal}
\end{equation}
The average of $G$ over one of the azimuths is exactly its large-distance form,
\begin{equation}
\frac{1}{2\pi}\int_{0}^{2\pi}G(\zeta,\zeta')\,\dd\phi'=-\frac{|w-w'|}{2},\label{eq:Gavg}
\end{equation}
because, for $w>w'$, $G=-\tfrac12(w-w')-\ln|1-e^{-(\zeta-\zeta')}|$ and the average of $\ln|1-\xi e^{i\theta}|$ over $\theta$ vanishes for $|\xi|<1$, by the mean-value property of harmonic functions.

This is the central structural fact. The conformal factor drops out of Poisson's equation, and two particles on the pseudosphere interact exactly as two particles on a flat cylinder. The curvature enters through the area element, and with it through the background. The background is uniform per unit of curved area, $-qn_{b}\,\dd A$, and since $\dd A=(a^{2}/w^{2})\,\dd w\,\dd\phi$, on the flat cylinder it has the non-uniform density $a^{2}n_{b}/w^{2}$, concentrated near the rim, Fig.~\ref{fig:system}(b). At every coupling, then, the plasma on the pseudosphere is a plasma on a flat cylinder with a non-uniform background, sampled with the area element $(a^{2}/w^{2})\,\dd w\,\dd\phi$.

\section{Background and neutrality}\label{sec:background}

With the area (\ref{eq:area}), the neutrality condition (\ref{eq:neutral0}) reads
\begin{equation}
N=\nu\Bigl(1-\frac1W\Bigr),\qquad \nu:=2\pi n_{b}a^{2}.\label{eq:neutral}
\end{equation}
The number $\nu$ is the number of particles the full, untruncated trumpet would hold at the same density, its area being $2\pi a^{2}$. In what follows $N$ and $\umax$ are the independent parameters, and $n_{b}$, hence $\nu$, is fixed by Eq.~(\ref{eq:neutral}).

\begin{figure}[t]
\centering
\includegraphics[width=\columnwidth]{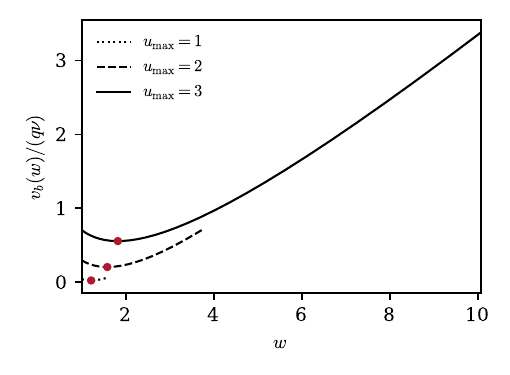}
\caption{The background potential (\ref{eq:vb}), in units of $q\nu$, on the band $1\le w\le W$ for $u_{\max}=1$, $2$ and $3$. The circles mark the minima at $w=2W/(W+1)$. The linear growth at large $w$ confines the particles; the term $-2\ln w$ is the trace of the curvature.}
\label{fig:vb}
\end{figure}

The background potential (\ref{eq:vbdef}) at a point of coordinate $w$ is reduced by the average (\ref{eq:Gavg}) to a single integral,
\begin{equation}
\int_{S}G\,\dd A'=2\pi a^{2}\int_{1}^{W}\Bigl(-\frac{|w-w'|}{2}\Bigr)\frac{\dd w'}{w'^{2}}.\label{eq:vbint}
\end{equation}
Splitting the interval at $w'=w$, the two pieces are elementary:
\begin{align}
\int_{1}^{w}\frac{w-w'}{w'^{2}}\,\dd w'&=w-1-\ln w,\nonumber\\
\int_{w}^{W}\frac{w'-w}{w'^{2}}\,\dd w'&=\ln W-\ln w+\frac{w}{W}-1.\nonumber
\end{align}
Adding them, with $\nu=2\pi n_{b}a^{2}$,
\begin{equation}
v_{b}(w)=\frac{q\nu}{2}\Bigl[\,w\Bigl(1+\frac1W\Bigr)-2\ln w+\ln W-2\Bigr].\label{eq:vb}
\end{equation}
The potential is confining: it grows linearly at large $w$, and its minimum, at $w=2W/(W+1)$, lies inside the band, Fig.~\ref{fig:vb}. The term $-2\ln w$, which comes from the weight $1/w'^{2}$ of the background, is the only trace of the curvature in it. Integrating $v_{b}$ against the background with the same weight, and using
\begin{multline}
\int_{1}^{W}\Bigl[w\Bigl(1+\frac1W\Bigr)-2\ln w+\ln W-2\Bigr]\frac{\dd w}{w^{2}}\\
=2\ln W+\frac{2\ln W}{W}+\frac4W-4,\nonumber
\end{multline}
the self-energy (\ref{eq:Ebbdef}) is
\begin{equation}
E_{bb}=-\frac{q^{2}\nu^{2}}{2}\Bigl[\ln W\Bigl(1+\frac1W\Bigr)+\frac2W-2\Bigr].\label{eq:Ebb}
\end{equation}
Thus, the potential energy (\ref{eq:U}) is now fully specified.

\section{Exact solution at $\Gamma=2$}\label{sec:exact}

\subsection{The Boltzmann factor as a squared determinant}

At $\Gamma=2$ the pair factor is $e^{-2G(\zeta_{i},\zeta_{j})}=|2\sinh((\zeta_{i}-\zeta_{j})/2)|^{2}$, and the one-body factor is, from Eq.~(\ref{eq:vb}) with $\beta q=2/q$,
\begin{equation}
e^{-\beta qv_{b}(w)}=\exp\Bigl\{-\nu\Bigl[w\Bigl(1+\frac1W\Bigr)-2\ln w+\ln W-2\Bigr]\Bigr\}.\label{eq:onebody}
\end{equation}
The pair factor becomes a Vandermonde determinant in the variables $e^{\zeta_{i}}$, as on the flat cylinder \cite{CFS1983}: since $2\sinh((\zeta_{i}-\zeta_{j})/2)=e^{-(\zeta_{i}+\zeta_{j})/2}(e^{\zeta_{i}}-e^{\zeta_{j}})$ and each $\zeta_{i}$ appears in $N-1$ pairs, the product of the pair factors is $|\det(e^{k\zeta_{i}})|^{2}\prod_{i}e^{-(N-1)w_{i}}$, with $i=1,\dots,N$ and $k=0,\dots,N-1$. The part of the exponent of particle $i$ that is linear in $w_{i}$ thus collects $-(N-1)w_{i}$ from the pairs and $-\nu(1+1/W)w_{i}$ from the background. By neutrality, $N=\nu-\nu/W$, so that
\begin{equation}
N-1+\nu\Bigl(1+\frac1W\Bigr)=2\nu-1.\label{eq:cancel}
\end{equation}
The position $W$ of the cut has disappeared from the one-body weight. It entered twice, through the background, which occupies only the truncated surface, and through $N$, which depends on $W$ by Eq.~(\ref{eq:neutral}); neutrality makes the two dependences cancel. Physically, the plasma on the truncated surface feels the same confinement as on the full trumpet, and only the domain of integration remembers the cut.

The background factor $e^{2\nu\ln w}=w^{2\nu}$ and the factor $w^{-2}$ of the area element combine, and the partition function (\ref{eq:ZNdef}) takes the form
\begin{equation}
Z_{N}=\frac{e^{-\nu N(\ln W-2)}}{N!}\!\int\!\prod_{i=1}^{N}\bigl[g(w_{i})\,\dd w_{i}\,\dd\phi_{i}\bigr]\bigl|\det\bigl(e^{k\zeta_{i}}\bigr)\bigr|^{2},\label{eq:ZN}
\end{equation}
with the one-body weight
\begin{equation}
g(w)=a^{2}\,w^{2\nu-2}\,e^{-(2\nu-1)w},\label{eq:g}
\end{equation}
the prefactor collecting the constant terms of the exponent (\ref{eq:onebody}). This is the form (\ref{eq:andreief}) with $\psi_{k}(\zeta)=e^{k\zeta}$ and $\dd\mu=g(w)\,\dd w\,\dd\phi$.

\subsection{Partition function and free energy}

The functions $e^{k\zeta}=e^{kw}e^{ik\phi}$ are orthogonal with respect to $\dd\mu$, the integral over the azimuth selecting equal angular momenta, and Andr\'eief's identity gives
\begin{equation}
Z_{N}=e^{-\nu N(\ln W-2)}\prod_{k=0}^{N-1}2\pi a^{2}I_{k},\label{eq:ZNprod}
\end{equation}
\begin{equation}
I_{k}:=\int_{1}^{W}w^{2\nu-2}\,e^{-(2\nu-2k-1)w}\,\dd w.\label{eq:Ikdef}
\end{equation}
With the decay rate $c_{k}:=2\nu-2k-1$ of orbital $k$, positive for all $k\le N-1$ because $N<\nu$, the substitution $t=c_{k}w$ expresses $I_{k}$ through the upper incomplete gamma function $\Gamma(s,x)=\int_{x}^{\infty}t^{s-1}e^{-t}\dd t$:
\begin{equation}
I_{k}=c_{k}^{-(2\nu-1)}\bigl[\Gamma(2\nu-1,c_{k})-\Gamma(2\nu-1,c_{k}W)\bigr].\label{eq:Ik}
\end{equation}
The norms are incomplete gamma functions because the surface has two boundaries; on the plane without walls the corresponding integrals run over $(0,\infty)$ and give complete gamma functions. Adding the background self-energy (\ref{eq:Ebb}), $\beta E_{bb}=\Gamma E_{bb}/q^{2}$ with $\Gamma=2$, the free energy (\ref{eq:Fdef}) of the configurational system is
\begin{multline}
\beta F=\nu N(\ln W-2)-\sum_{k=0}^{N-1}\ln\bigl(2\pi a^{2}I_{k}\bigr)\\
-\nu^{2}\Bigl[\ln W\Bigl(1+\frac1W\Bigr)+\frac2W-2\Bigr],\label{eq:F}
\end{multline}
exact for every $N$ and $\umax$. As an independent check, for $N=2$ and $\umax=2$ the partition function obtained by direct numerical integration of Eq.~(\ref{eq:ZN}) agrees with the product (\ref{eq:ZNprod}) to eighteen digits.

\subsection{Density, correlations and the orbital picture}

By Sec.~\ref{sec:whygamma2}, the one-body density with respect to the flat measure $\dd w\,\dd\phi$ is $g(w)\sum_{k}|\psi_{k}|^{2}/h_{k}$, and dividing by the conformal factor $a^{2}/w^{2}$ gives the number of particles per unit curved area,
\begin{equation}
n(w)=\frac{w^{2\nu}}{2\pi a^{2}}\sum_{k=0}^{N-1}\frac{e^{-(2\nu-2k-1)w}}{I_{k}},\qquad \int_{S}n\,\dd A=N,\label{eq:density}
\end{equation}
the normalization following term by term from Eq.~(\ref{eq:Ikdef}). The truncated pair correlation is $-|K_{N}(\zeta,\zeta')|^{2}$, with the kernel
\begin{equation}
K_{N}(\zeta,\zeta')=\frac{(ww')^{\nu}}{2\pi a^{2}}\,e^{-(\nu-\frac12)(w+w')}\sum_{k=0}^{N-1}\frac{e^{k(\zeta+\bar\zeta')}}{I_{k}},\label{eq:kernel}
\end{equation}
and every higher correlation is a determinant built from $K_{N}$.

\begin{figure}[t]
\centering
\includegraphics[width=\columnwidth]{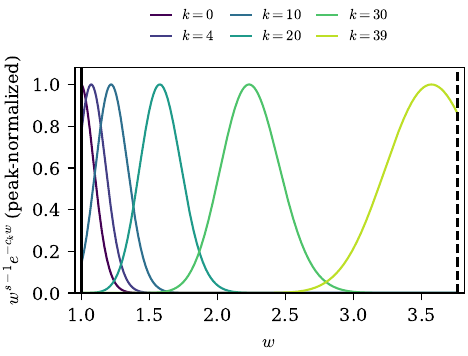}
\caption{The orbital ladder: the one-body weights $w^{s-1}e^{-c_{k}w}$ of the integrals (\ref{eq:Ikdef}), $s=2\nu-1$, $c_{k}=s-2k$, for $N=40$, $\umax=2$ and six orbitals, each normalized to its peak. The vertical lines are the rim (solid) and the cut (dashed). The peak of orbital $k$ moves from the rim ($k=0$) to the cut ($k=N-1$); only the orbitals whose bumps are cut by a wall feel the boundaries.}
\label{fig:orbitals}
\end{figure}

Each term of Eq.~(\ref{eq:density}) has a simple meaning. The function $w^{2\nu}e^{-c_{k}w}$ is a bump peaked at $w_{k}^{*}=2\nu/c_{k}$, of width of order $\sqrt{2\nu}/c_{k}$, Fig.~\ref{fig:orbitals}. As $k$ runs from $0$ to $N-1$ the peak moves from the rim, $w_{0}^{*}\simeq1$, to the cut, since $c_{N-1}=2\nu/W+1$ by neutrality. The $N$ orbitals fill the band in order, one per unit of angular momentum, as the Landau-level orbitals fill a disc in the plane. Orbitals whose bumps lie well inside $(1,W)$ do not feel the walls; only the first few orbitals at the rim and the last few at the cut do. This picture organizes everything that follows.

\section{The plasma at finite $N$}\label{sec:finiteN}

\subsection{Two boundary layers}

\begin{figure}[t]
\centering
\includegraphics[width=\columnwidth]{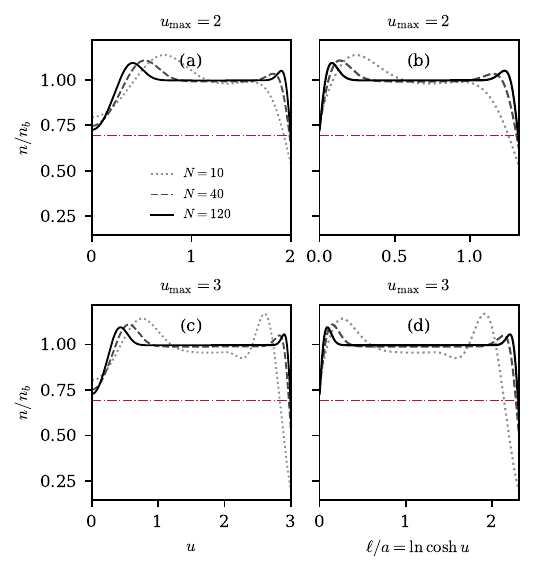}
\caption{Exact density (\ref{eq:density}) in units of the background density for $\umax=2$ (top) and $\umax=3$ (bottom) and $N=10$, $40$, $120$, against the meridian parameter $u$ in (a) and (c) and against the proper length $\ell=a\ln\cosh u$ along the meridian in (b) and (d). The dash-dotted line is the limiting contact value $\ln2$ of Proposition~\ref{prop:contact}. In the parameter $u$ the layer at the rim is stretched, because $\dd\ell/\dd u=a\tanh u$ vanishes there; in proper length the two layers have the same thickness. The two walls are depleted differently at finite $N$, more deeply at the cut, whose circumference is $W$ times shorter than that of the rim.}
\label{fig:density}
\end{figure}

\begin{figure*}[!t]
\centering
\includegraphics[width=\textwidth]{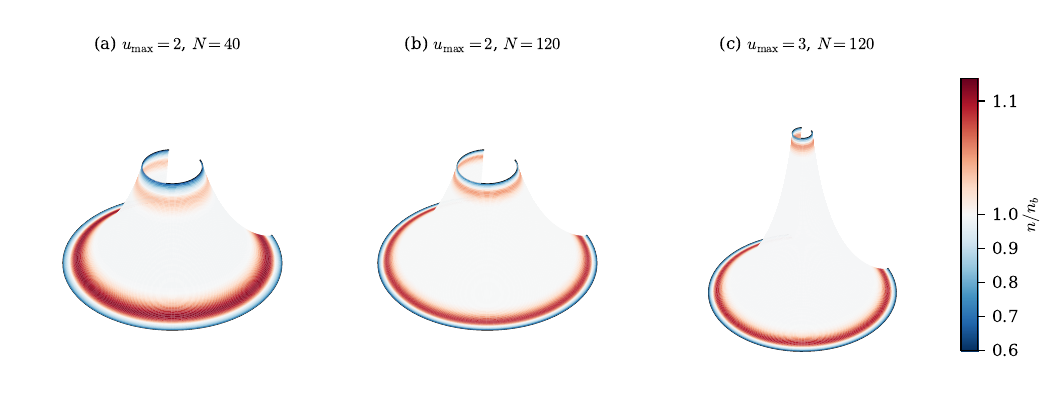}
\caption{Exact density (\ref{eq:density}) at $\Gamma=2$ drawn on the truncated pseudosphere (\ref{eq:embed}), for (a) $\umax=2$, $N=40$, (b) $\umax=2$, $N=120$ and (c) $\umax=3$, $N=120$. The colour gives $n/n_{b}$ on a scale centred on the bulk value (white), with depletion in blue and excess in red; a sector of the surface is removed to show its shape. The wide circle at the bottom is the rim $u=0$ and the narrow one at the top the cut $u=\umax$.}
\label{fig:density3d}
\end{figure*}

Figure~\ref{fig:density} shows the density (\ref{eq:density}) for $\umax=2$ and $3$. In panels (a) and (c), where it is plotted against the parameter $u$, the profile is much steeper at the cut than at the rim, where it starts with zero slope. This is a property of the coordinate and not of the plasma. The proper length along the meridian is $\dd\ell=a\tanh u\,\dd u$ by Eq.~(\ref{eq:metric}), so that $\dd n/\dd u=a\tanh u\,\dd n/\dd\ell$ vanishes at $u=0$ whatever the slope $\dd n/\dd\ell$ is. A layer of proper thickness $\delta\ll a$ occupies the interval $\Delta u\simeq\sqrt{2\delta/a}$ at the rim and $\Delta u\simeq\delta/(a\tanh\umax)$ at the cut; for $\umax=2$ and $N=120$, where the interparticle spacing is $n_{b}^{-1/2}=0.196\,a$, one spacing is $\Delta u=0.65$ at the rim and $\Delta u=0.21$ at the cut. Panels (b) and (d) show the same profiles against the proper length $\ell=a\ln\cosh u$ of Eq.~(\ref{eq:ell}), in which the two layers have the same thickness and the same structure. In the interior the density equals $n_{b}$ to high accuracy already at $N=40$: the bulk is neutral and uniform, as on any surface of constant curvature. Near each wall the density is depleted, overshoots at about one interparticle spacing, and relaxes to $n_{b}$, which is the structure of the flat plasma against a hard wall \cite{Jancovici1982}. What differs is the depth of the depletion. At $\umax=3$ and $N=120$ the density is $0.73\,n_{b}$ at the rim and $0.62\,n_{b}$ at the cut, the latter still rising with $N$ ($0.52\,n_{b}$ at $N=40$). Both walls are hard walls and horocycles of the same curvature, so no local property distinguishes them; the difference is a global effect. The curvature fixes the circumference $2\pi a/w$ of each parallel, and a particle at the rim has $W$ times more neighbours along its wall than a particle at the cut. The correlation hole around a particle, of the size of the interparticle spacing, therefore contains different numbers of particles at the two walls as long as the spacing is not negligible against the circumference of the cut, that is, at finite $N$.

Figure~\ref{fig:density3d} shows the same densities on the surface itself. Each wall carries a depleted strip at contact followed by a ring in which the density exceeds its bulk value, and the two layers have the same thickness, of the order of the interparticle spacing; the layer at the cut lies on a circle shorter by the factor $W$ and holds correspondingly fewer particles.

\subsection{Thin band and full trumpet}

For $\umax\to0$ the band becomes narrow. Its circumference is $2\pi a$ at both walls to leading order, its proper height is $\ell\simeq a\umax^{2}/2$ by Eq.~(\ref{eq:ell}), and its area (\ref{eq:area}) is $A\simeq\pi a^{2}\umax^{2}$, the product of the two. In the conformal coordinate $W-1\simeq\umax^{2}/2$, the factor $w^{2\nu-2}$ of the weight is $1+O(\umax^{2})$, and the orbitals reduce to the exponentials $e^{-c_{k}(w-1)}$ of the plasma on a flat cylinder, the semiperiodic strip of Choquard, Forrester and Smith \cite{CFS1983}. The curvature is invisible on a band whose height is small compared with $a$.

For $\umax\to\infty$ the cut recedes down the cusp, the area tends to $2\pi a^{2}$ and $N\to\nu$; the second incomplete gamma function in Eq.~(\ref{eq:Ik}) vanishes. The last orbital, $k=\nu-1$, has $c_{k}=1$ and the weight $w^{2\nu-2}e^{-w}$, peaked at $w=2\nu-2$: the outermost particles reach down the cusp to a circumference of order $2\pi a/\nu$, comparable with the interparticle spacing. Beyond this scale the density is exponentially small, and the cusp acts as a natural cut-off rather than as a boundary.

\section{The thermodynamic limit}\label{sec:largeN}

\subsection{Fixed density and fixed size}

The thermodynamic limit takes $N\to\infty$ at fixed background density $n_{b}$, with the area growing as $N$. On a surface of fixed radius of curvature this is impossible, the area (\ref{eq:area}) being bounded by $2\pi a^{2}$. The limit is therefore taken as on the sphere, whose radius grows with $N$ at fixed density \cite{Caillol1981}: the shape $\umax$ is fixed and the radius of curvature grows according to the neutrality relation (\ref{eq:neutral}),
\begin{equation}
a^{2}=\frac{N}{2\pi\alpha\,n_{b}},\qquad \alpha:=1-\frac1W,\qquad \nu=\frac N\alpha.\label{eq:fixedshape}
\end{equation}
The surfaces of this sequence differ by a change of scale. The interparticle spacing $n_{b}^{-1/2}$ is fixed, the Gaussian curvature $-1/a^{2}$ and the geodesic curvature $1/a$ of the walls tend to zero, and the surface becomes flat on the scale of the spacing, while the two curvature integrals of Eq.~(\ref{eq:GB}), $-2\pi\alpha$ and $+2\pi\alpha$, do not depend on $a$. On the sphere, likewise, the curvature $1/R^{2}$ tends to zero and its integral remains $4\pi$. In such a limit the local curvature disappears from the bulk properties, as noted by Jancovici and T\'ellez \cite{JancoviciTellez1998}, and the geometry enters only through the integrated curvatures, in the boundary and constant terms of the free energy and in the coefficient of its logarithmic term \cite{JMP1994}. This is the situation to which the structure (\ref{eq:Fstructure}) refers, with $L_{b}\propto a\propto\sqrt N$.

The same limit can be described on a surface of fixed size. In the conformal coordinates neither the potential (\ref{eq:G}) nor the product $n_{b}\,\dd A=(\nu/2\pi)\,\dd w\,\dd\phi/w^{2}$ contains $a$, which enters $Z_{N}$ only through the area elements of the particles, as the factor $a^{2N}$. Hence $\beta F$ depends on $a$ only through the term $-N\ln a^{2}$ at fixed $N$ and $\umax$, and letting $N$ grow at fixed $a$, the density then growing as $N$, gives the same expansion as the limit (\ref{eq:fixedshape}) up to this explicit term. This is the absence of an intrinsic length in the logarithmic interaction, and it is the form in which the large-$N$ limit of random normal matrices is usually taken, $N$ growing in a fixed domain \cite{ZabrodinWiegmann2006,Klevtsov2014}. One more explicit term arises from the zero of the potential. The potential (\ref{eq:G}) measures distances in units of $a$ (Sec.~\ref{sec:surface}), a unit that grows along the sequence (\ref{eq:fixedshape}). With a fixed reference length $L_{0}$ the potential is $G-\ln(a/L_{0})$, and the three terms of Eq.~(\ref{eq:U}) change by $-\ln(a/L_{0})$ times $q^{2}N(N-1)/2$, $-q^{2}N^{2}$ and $q^{2}N^{2}/2$, respectively, that is, by $(q^{2}N/2)\ln(a/L_{0})$ in total. At $\Gamma=2$, the free energy for the reference length $L_{0}$ is therefore
\begin{equation}
\beta F_{L_{0}}=\beta F+N\ln\frac{a}{L_{0}}.\label{eq:FL0}
\end{equation}

In the following subsections the expansion is derived in the variable $N$ at fixed $a$, in which the formulas are simplest. The microscopic length $n_{b}^{-1/2}$ then shrinks as $N^{-1/2}$ against the geometric length $a$, and every finite-size effect is organized by the ratio $a\sqrt{n_{b}}\propto\sqrt N$. The boundary layers, one interparticle spacing thick, contain $O(\sqrt N)$ orbitals; the boundary free energy, a tension times a length, is proportional to $\sqrt{n_{b}}\propto\sqrt N$; the bulk free energy per particle at $\Gamma=2$, which is $\tfrac12\ln n_{b}$ plus a constant \cite{Jancovici1981,AlastueyJancovici1981}, produces $\tfrac12N\ln N$; and the universal term, proportional to $\ln(a\sqrt{n_{b}})$, becomes proportional to $\tfrac12\ln N$. The expected form is therefore
\begin{equation}
\beta F=\tfrac12N\ln N+A_{2}N+B\sqrt N+C\ln N+D+\dots,\label{eq:expected}
\end{equation}
with $C$ the universal coefficient. Each coefficient can be derived from the exact solution, and the result is then written at fixed density in Sec.~\ref{sec:reading}.

\subsection{Structure of the exact free energy}

With $s:=2\nu-1=2N/\alpha-1$, the free energy (\ref{eq:F}) is $\beta F=T_{1}+T_{2}+T_{3}+T_{4}$, where
\begin{align}
T_{1}&=\frac{N^{2}}{\alpha}(\ln W-2),\qquad T_{2}=-N\ln(2\pi a^{2}),\label{eq:T12}\\
T_{3}&=-\sum_{k=0}^{N-1}\ln I_{k},\label{eq:T3}\\
T_{4}&=-\frac{N^{2}}{\alpha^{2}}\Bigl[\ln W\Bigl(1+\frac1W\Bigr)+\frac2W-2\Bigr].\label{eq:T4}
\end{align}
$T_{1}$ and $T_{4}$ are of order $N^{2}$: they are the mean-field energies of the background, and must cancel against the order-$N^{2}$ part of $T_{3}$, since a neutral system cannot have a free energy of order $N^{2}$. With the regularized incomplete gamma function $Q(s,x)=\Gamma(s,x)/\Gamma(s)$,
\begin{equation}
\ln I_{k}=\ln\Gamma(s)-s\ln c_{k}+\ln b_{k},\qquad c_{k}=s-2k,\label{eq:lnIk}
\end{equation}
where
\begin{equation}
b_{k}:=Q(s,c_{k})-Q(s,c_{k}W)\label{eq:bk}
\end{equation}
is the fraction of the gamma density of orbital $k$ that lies between the walls. The three pieces of Eq.~(\ref{eq:lnIk}) are treated in turn.

\subsection{The Stirling piece}

With $S:=2N/\alpha=s+1$, Stirling's series for $\ln\Gamma(s)$ and the expansion $\ln s=\ln S-1/S-1/(2S^{2})+O(S^{-3})$ give
\begin{equation}
\ln\Gamma(s)=S\ln S-S-\tfrac32\ln S+\tfrac12\ln2\pi+\frac{13}{12S}+O(S^{-2}),\label{eq:stirling}
\end{equation}
and, multiplying by $-N$,
\begin{multline}
-N\ln\Gamma(s)=-\frac{2N^{2}}{\alpha}\ln\frac{2N}{\alpha}+\frac{2N^{2}}{\alpha}+\frac{3N}{2}\ln\frac{2N}{\alpha}\\
-\frac N2\ln2\pi-\frac{13\alpha}{24}+O(N^{-1}).\label{eq:stirlingN}
\end{multline}

\subsection{The exact sum of $\ln c_{k}$}

Since $c_{k}=2(\nu-\tfrac12-k)$, the product of the $c_{k}$ is a ratio of gamma functions. Using $\nu-N=\nu/W$ from neutrality,
\begin{equation}
\sum_{k=0}^{N-1}\ln c_{k}=N\ln2+\ln\Gamma\bigl(\nu+\tfrac12\bigr)-\ln\Gamma\bigl(\tfrac{\nu}{W}+\tfrac12\bigr).\label{eq:sumlnc}
\end{equation}
The expansion
\begin{equation}
\ln\Gamma\bigl(x+\tfrac12\bigr)=x\ln x-x+\tfrac12\ln2\pi-\frac{1}{24x}+O(x^{-3})\label{eq:lngammahalf}
\end{equation}
contains no term in $\ln x$; this fact will turn out to be the origin of the absence of $\ln N$. With $\nu=N/\alpha$ it gives
\begin{multline}
\ln\Gamma\bigl(\nu+\tfrac12\bigr)-\ln\Gamma\bigl(\tfrac{\nu}{W}+\tfrac12\bigr)=\nu\alpha\ln\nu+\frac{\nu}{W}\ln W\\
-\nu\alpha+\frac{W-1}{24\nu}+O(\nu^{-3}),\label{eq:lngammadiff}
\end{multline}
and, multiplying by $s=2\nu-1$,
\begin{multline}
s\sum_{k}\ln c_{k}=\frac{2N^{2}}{\alpha}\ln\frac{2N}{\alpha}+\frac{2N^{2}\ln W}{\alpha^{2}W}-\frac{2N^{2}}{\alpha}-N\ln\frac{2N}{\alpha}\\
-\frac{N\ln W}{\alpha W}+N+\frac{W-1}{12}+O(N^{-1}).\label{eq:sumlncN}
\end{multline}

\subsection{The bracket: two boundary layers}

The integrand of $I_{k}$, $w^{s-1}e^{-c_{k}w}$, is a gamma density in $w$ peaked at $w_{k}^{*}=(s-1)/c_{k}$, of width of order $\sqrt s/c_{k}$. For an orbital whose peak lies inside $(1,W)$ by more than a few widths, $b_{k}$ is $1$ up to exponentially small terms: the interior orbitals contribute nothing at any algebraic order. The peak reaches the rim at $k=0$ and the cut at $k=N-1$ up to $O(1)$, and the two ends of the orbital ladder form two boundary layers of $O(\sqrt s)$ orbitals each. In each layer Temme's uniform expansion of the incomplete gamma function \cite{Temme1975,DLMF} applies,
\begin{equation}
Q(s,s\lambda)=\tfrac12\erfc\bigl(\eta\sqrt{s/2}\bigr)+O(s^{-1/2}),\label{eq:temme}
\end{equation}
with $\eta=\operatorname{sign}(\lambda-1)\sqrt{2(\lambda-1-\ln\lambda)}=\lambda-1+O((\lambda-1)^{2})$.

At the rim, $k\ll N$, one has $\lambda_{k}=c_{k}/s=1-2k/s$ and $\eta_{k}\sqrt{s/2}=-t_{k}$ with $t_{k}=k\sqrt{2/s}\,[1+O(k/s)]$, while $Q(s,c_{k}W)$ is negligible. With $u(t):=1-\tfrac12\erfc t$,
\begin{equation}
-\sum_{k\in\mathrm{rim}}\ln b_{k}=\sum_{k\ge0}g(t_{k})+O(1),\qquad g(t):=-\ln u(t).\label{eq:rimsum}
\end{equation}
The function $g$ decreases from $g(0)=\ln2$ to zero faster than any power. For a function $f$ of this kind, sampled with a step $h$ from an offset $\tau\ge0$, the Euler--Maclaurin formula reads
\begin{equation}
\sum_{j\ge0}f(\tau+jh)=\frac1h\int_{\tau}^{\infty}f(t)\,\dd t+\frac12f(\tau)+O(h).\label{eq:EM}
\end{equation}
With $f=g$, $\tau=0$ and $h=\sqrt{2/s}$, the leading term of the sum (\ref{eq:rimsum}) is $\sqrt{s/2}\,J$, with the single quadrature
\begin{equation}
J:=-\int_{0}^{\infty}\ln\Bigl[1-\tfrac12\erfc t\Bigr]\dd t=0.337668477\ldots\label{eq:J}
\end{equation}
At the cut, $k=N-1-j$ with $j\ll N$, one has $c_{k}W-s=W+1+2Wj$. With
\begin{equation}
t_{j}:=\frac{c_{k}W-s}{\sqrt{2s}}=t_{0}+jW\sqrt{2/s},\qquad t_{0}:=\frac{W+1}{\sqrt{2s}},\label{eq:tj}
\end{equation}
Eq.~(\ref{eq:temme}) gives $1-Q(s,c_{k}W)=u(t_{j})+O(s^{-1/2})$, while $Q(s,c_{k})$ is exponentially close to $1$. This layer has the same profile $g$, sampled with a step $W$ times larger, and by Eq.~(\ref{eq:EM}) its leading contribution is $\sqrt{s/2}\,J/W$. The larger step is the shorter circumference: the cut holds $W$ times fewer particles per orbital, and its layer, of the same thickness in units of the spacing, contains $W$ times fewer orbitals. With $\sqrt{s/2}=\sqrt{N/\alpha}+O(N^{-1/2})$,
\begin{equation}
-\sum_{k=0}^{N-1}\ln b_{k}=J\Bigl(1+\frac1W\Bigr)\sqrt{\frac N\alpha}+D_{\mathrm{rim}}+D_{\mathrm{cut}}+O(N^{-1/2}).\label{eq:bracket}
\end{equation}

The constants $D_{\mathrm{rim}}$ and $D_{\mathrm{cut}}$ require the next term of Temme's expansion,
\begin{multline}
Q(s,s\lambda)=\tfrac12\erfc\bigl(\eta\sqrt{s/2}\bigr)\\
+\frac{e^{-s\eta^{2}/2}}{\sqrt{2\pi s}}\Bigl[\frac{1}{\lambda-1}-\frac1\eta\Bigr]+O(s^{-3/2}),\label{eq:temme2}
\end{multline}
together with $\eta=(\lambda-1)[1-\tfrac13(\lambda-1)+O((\lambda-1)^{2})]$, for which the square bracket tends to $-\tfrac13$ as $\lambda\to1$. In the rim layer, where $\lambda-1=-t\sqrt{2/s}$,
\begin{equation}
Q(s,c_{k})=u(t_{k})+\frac{e^{-t_{k}^{2}}(2t_{k}^{2}-1)}{3\sqrt{2\pi s}}+O(s^{-1}),\label{eq:temmerim}
\end{equation}
and in the cut layer the correction has the opposite sign,
\begin{equation}
1-Q(s,c_{k}W)=u(t_{j})-\frac{e^{-t_{j}^{2}}(2t_{j}^{2}-1)}{3\sqrt{2\pi s}}+O(s^{-1}),\label{eq:temmecut}
\end{equation}
because the two layers lie on opposite sides of the peak. Write $\delta(t):=e^{-t^{2}}(2t^{2}-1)/(3\sqrt{2\pi})$ for the correction. At the rim $\ln b_{k}=\ln u(t_{k})+s^{-1/2}\delta(t_{k})/u(t_{k})+O(s^{-1})$, so that
\begin{equation}
-\sum_{k\in\mathrm{rim}}\ln b_{k}=\sum_{k\ge0}g(t_{k})-\frac{1}{\sqrt s}\sum_{k\ge0}\frac{\delta(t_{k})}{u(t_{k})}+O(s^{-1/2}).\label{eq:rimtwo}
\end{equation}
By Eq.~(\ref{eq:EM}) with $\tau=0$ and $h=\sqrt{2/s}$, the first sum is $\sqrt{s/2}\,J+\tfrac12\ln2+O(s^{-1/2})$. The second is $s^{-1/2}\sqrt{s/2}\int_{0}^{\infty}(\delta/u)\,\dd t+O(s^{-1/2})$; with the second quadrature
\begin{equation}
M:=\int_{0}^{\infty}\frac{e^{-t^{2}}(1-2t^{2})}{u(t)}\,\dd t=0.242910230\ldots\label{eq:M}
\end{equation}
one has $\int_{0}^{\infty}(\delta/u)\,\dd t=-M/(3\sqrt{2\pi})$, so that the second sum equals $-M/(6\sqrt\pi)$. Hence $D_{\mathrm{rim}}=\tfrac12\ln2+M/(6\sqrt\pi)$. At the cut the correction enters with the opposite sign, and the samples start at the offset $t_{0}$ of Eq.~(\ref{eq:tj}) with the step $W\sqrt{2/s}$:
\begin{equation}
-\sum_{k\in\mathrm{cut}}\ln b_{k}=\sum_{j\ge0}g(t_{j})+\frac{1}{\sqrt s}\sum_{j\ge0}\frac{\delta(t_{j})}{u(t_{j})}+O(s^{-1/2}).\label{eq:cuttwo}
\end{equation}
In the first sum, Eq.~(\ref{eq:EM}) with $\tau=t_{0}$, $h=W\sqrt{2/s}$, $\int_{t_{0}}^{\infty}g\,\dd t=J-t_{0}\ln2+O(s^{-1})$ and $t_{0}/(W\sqrt{2/s})=(W+1)/(2W)$ gives
\begin{equation}
\sum_{j\ge0}g(t_{j})=\sqrt{\frac s2}\,\frac JW+\ln2\Bigl[\frac12-\frac{W+1}{2W}\Bigr]+O(s^{-1/2}),\nonumber
\end{equation}
whose constant is $-\ln2/(2W)$; the second sum gives $-M/(6\sqrt\pi\,W)$ by the same steps as at the rim. Altogether,
\begin{equation}
D_{\mathrm{rim}}=\frac{\ln2}{2}+\frac{M}{6\sqrt\pi},\qquad D_{\mathrm{cut}}=-\frac{1}{W}\Bigl(\frac{\ln2}{2}+\frac{M}{6\sqrt\pi}\Bigr).\label{eq:DM}
\end{equation}

\subsection{Assembly}

\begin{proposition}\label{prop:F}
For $N\to\infty$ at fixed $\umax$ and fixed $a$,
\begin{equation}
\beta F=\frac N2\ln N+A_{2}N+B\sqrt N+D+O(N^{-1/2}),\label{eq:expansion}
\end{equation}
with no term in $\ln N$, and with
\begin{multline}
A_{2}=1+\frac{\ln2}{2}-\frac12\ln\Bigl(1-\frac1W\Bigr)-\frac32\ln2\pi\\
-\frac{\ln W}{W-1}-\ln a^{2},\label{eq:A2}
\end{multline}
\begin{equation}
B=J\,\frac{1+1/W}{\sqrt{1-1/W}},\label{eq:B}
\end{equation}
\begin{equation}
D=-\frac{13\alpha}{24}+\frac{W-1}{12}+\alpha\Bigl[\frac{\ln2}{2}+\frac{M}{6\sqrt\pi}\Bigr],\label{eq:D}
\end{equation}
$J$ and $M$ being given by Eqs.~(\ref{eq:J}) and (\ref{eq:M}).
\end{proposition}

\begin{proof}
Equations (\ref{eq:stirlingN}), (\ref{eq:sumlncN}) and (\ref{eq:bracket}) are collected into $T_{3}$ through Eq.~(\ref{eq:lnIk}), and $T_{1}$, $T_{2}$, $T_{4}$ are added. The $N^{2}\ln N$ terms cancel between Eqs.~(\ref{eq:stirlingN}) and (\ref{eq:sumlncN}). The $N^{2}$ terms add up to $N^{2}/\alpha^{2}$ times
\begin{equation}
\alpha(\ln W-2)+\frac{2\ln W}{W}-\ln W-\frac{\ln W}{W}-\frac2W+2=0,\label{eq:N2cancel}
\end{equation}
by $\alpha\ln W-\ln W=-(\ln W)/W$ and $2-2\alpha=2/W$; this is the cancellation of the mean-field energies. The $N\ln N$ terms are $\tfrac32N\ln N-N\ln N=\tfrac12N\ln N$. The terms linear in $N$ are
\begin{equation}
\frac{3N}{2}\ln\frac2\alpha-\frac N2\ln2\pi-N\ln\frac2\alpha+N-\frac{N\ln W}{\alpha W}-N\ln(2\pi a^{2}),\nonumber
\end{equation}
which is $A_{2}N$ since $\alpha W=W-1$. The $\sqrt N$ term is that of Eq.~(\ref{eq:bracket}). The constant collects $-13\alpha/24$ from Eq.~(\ref{eq:stirlingN}), $(W-1)/12$ from Eq.~(\ref{eq:sumlncN}), and $D_{\mathrm{rim}}+D_{\mathrm{cut}}=\alpha[\tfrac12\ln2+M/(6\sqrt\pi)]$. No piece produces $\ln N$: the Stirling piece has $\tfrac32\ln S$ only multiplied by $N$, the sum (\ref{eq:sumlnc}) has none because of Eq.~(\ref{eq:lngammahalf}), and the layers give $\sqrt N$ and constants.
\end{proof}

The coefficients $\tfrac12$ and $A_{2}$ follow from Stirling's series and the exact identity (\ref{eq:sumlnc}) alone. The coefficient $B$, the constant $D$ and the absence of a $\ln N$ term rest in addition on the two-term expansion (\ref{eq:temme2}) and on the Euler--Maclaurin formula (\ref{eq:EM}); their remainders are of order $N^{-1/2}$ in the derivation but are not bounded rigorously here, and the comparison with the exact free energy that follows is what tests them.

\subsection{Comparison with the exact free energy}\label{sec:validation}

The exact free energy (\ref{eq:F}) allows each coefficient of Proposition~\ref{prop:F} to be compared with the finite-$N$ solution. It is evaluated in the form $T_{1}+T_{4}-N\ln2\pi-N\ln\Gamma(s)+s\sum_{k}\ln c_{k}-\sum_{k}\ln b_{k}$ of Eqs.~(\ref{eq:T12})--(\ref{eq:bk}): the terms of order $N^{2}$, which cancel against each other, in 40-digit arithmetic with $\sum_{k}\ln c_{k}$ from the exact identity (\ref{eq:sumlnc}), and the brackets $b_{k}$, which are of order one, in double precision. Two comparisons are made. The first solves exactly the seven linear equations $\beta F(N)=\sum_{n}x_{n}\varphi_{n}(N)$ at $N=16\times2^{m}$, $m=0,\dots,6$, for the coefficients $x_{n}$ of the functions $N\ln N$, $N$, $\sqrt N$, $\ln N$, $1$, $N^{-1/2}$ and $N^{-1}$. At $\umax=2$ it returns $0.50000$, $-1.73545$, $0.49894$ and $-0.00073$ for the first four, against the derived $\tfrac12$, $A_{2}=-1.73545$, $B=0.49883$ and $C=0$. At $\umax=1$ and $3$ the higher-order corrections are larger and this elimination is less sharp, which is why the second test is more informative.

\begin{figure}[t]
\centering
\includegraphics[width=\columnwidth]{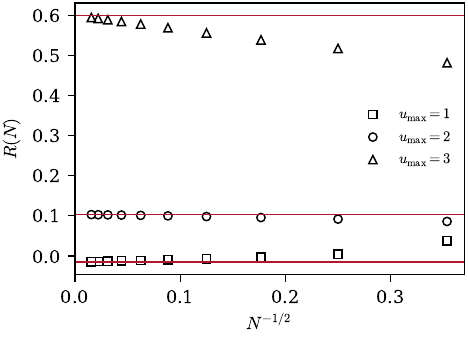}
\caption{The residual $R(N)$ of Eq.~(\ref{eq:residual}), from the exact free energy (\ref{eq:F}) with $a=1$, against $N^{-1/2}$ for $N=8,16,\dots,4096$ and three shapes. The horizontal lines are the derived constants $D$ of Eq.~(\ref{eq:D}). The approach is linear in $N^{-1/2}$, with no constant increment per doubling of $N$; Table~\ref{tab:residual} gives the numbers.}
\label{fig:residual}
\end{figure}

\begin{table}[t]
\caption{Test of Proposition~\ref{prop:F} against the exact free energy. $D$ is the derived constant (\ref{eq:D}); $R(4096)$ is the exact residual (\ref{eq:residual}); $R_{\infty}=[\sqrt2R(4096)-R(2048)]/(\sqrt2-1)$ removes the $N^{-1/2}$ correction; $\hat C$ is the estimate (\ref{eq:Chat}) of a $\ln N$ coefficient from $N=1024$, $2048$, $4096$.}
\label{tab:residual}
\begin{ruledtabular}
\begin{tabular}{ccccc}
$\umax$ & $D$ & $R(4096)$ & $R_{\infty}$ & $\hat C$\\
\colrule
1 & $-0.015367$ & $-0.014310$ & $-0.015364$ & $-4{\times}10^{-6}$\\
2 & $\phantom{-}0.103716$ & $\phantom{-}0.103112$ & $\phantom{-}0.103732$ & $-2{\times}10^{-5}$\\
3 & $\phantom{-}0.600496$ & $\phantom{-}0.595070$ & $\phantom{-}0.600523$ & $-3{\times}10^{-5}$\\
\end{tabular}
\end{ruledtabular}
\end{table}

The second test uses the residual
\begin{equation}
R(N):=\beta F-\frac N2\ln N-A_{2}N-B\sqrt N,\label{eq:residual}
\end{equation}
which must tend to $D$. A term $C\ln N$ would add $C\ln2$ to each increment $\delta_{m}=R(2^{m+1}N_{0})-R(2^{m}N_{0})$, while an $N^{-1/2}$ correction multiplies the increments by $2^{-1/2}$ at each doubling. The combination
\begin{equation}
\hat C=\frac{\delta_{m+1}-2^{-1/2}\delta_{m}}{(1-2^{-1/2})\ln2}\label{eq:Chat}
\end{equation}
therefore removes the $N^{-1/2}$ correction and gives an estimate of $C$. Figure~\ref{fig:residual} and Table~\ref{tab:residual} show the result for $N$ up to $4096$. The residual approaches $D$ linearly in $N^{-1/2}$; after removal of that correction it agrees with the derived $D$ to four or five digits for the three shapes; and the estimate $\hat C$ of a possible $\ln N$ coefficient is of order $10^{-5}$ at the largest sizes, decreasing by about a factor of two at each doubling of $N$, as a residual correction of order $N^{-1}$ would. The exact free energy is thus consistent with the expansion of Proposition~\ref{prop:F}, which has no logarithmic term.

The expansion can also be compared with the general cylinder result of Shen and Forrester \cite{ShenForrester2026}. They consider $N$ charges at $\Gamma=2$ on a cylinder with a one-body potential depending only on the axial coordinate, and obtain the large-$N$ expansion of the free energy through the constant term for soft walls and for hard walls at the edges of the droplet or inside it, with no term in $\ln N$ in any case. To write the present system in their form, put $z=e^{\zeta}=e^{w+i\phi}$ and $r=|z|=e^{w}$, so that $|2\sinh\tfrac{\zeta-\zeta'}{2}|^{2}=|z-z'|^{2}/(rr')$, and collect the one-body factors of the Boltzmann weight: the area element $(a^{2}/w^{2})\,\dd w\,\dd\phi=\dd^{2}z/(r^{2}\ln^{2}r)$ at $a=1$, the background factor $e^{-2v_{b}}=r^{-\nu(1+1/W)}(\ln r)^{2\nu}e^{-\nu(\ln W-2)}$ from Eq.~(\ref{eq:vb}), and the factor $r^{-(N-1)}$ left by the pair potential. With $N=\nu(1-1/W)$ the weight becomes, up to a constant,
\begin{align}
&\prod_{i}e^{-Nq(r_{i})}\,r_{i}^{-1}\,e^{a(r_{i})}\,\dd^{2}z_{i}\prod_{i<j}|z_{i}-z_{j}|^{2},\label{eq:SFform}\\
&q(r)=A\,(\ln r-\ln\ln r),\qquad A=\frac{2W}{W-1},\nonumber\\
&a(r)=-2\ln\ln r,\nonumber
\end{align}
which is the form of their Eqs.~(1.14)--(1.15), the factor $e^{a}$ being the linear statistic treated by their Theorem~1. The droplet of $q$, bounded by the roots of $rq'(r)=0$ and $rq'(r)=2$, is $e\le r\le e^{W}$, that is $1\le w\le W$: the rim and the cut of the pseudosphere are hard walls at the two edges of the droplet, the case of their Theorem~2 with both walls critical. The identity $rq'(r)/2=(W/(W-1))(1-1/\ln r)$ shows that their variable $\tau$ is the fraction of the orbitals below a given $w$, and two of their universal constants are the quadratures of Sec.~\ref{sec:largeN}: the substitution $x=-t$ in their definitions gives $\gamma^{\rm in}=J/\sqrt2$ and $\tilde\alpha^{\rm in}=M/(6\sqrt\pi)$. Evaluating their Theorems~1 and 2 for the potential (\ref{eq:SFform}), in which the quadratic term of Theorem~1 vanishes identically for this $a$, and subtracting the result from the exact $\ln Z_{N}$ gives $-0.0011$, $+0.0006$ and $+0.0054$ at $N=4096$ for $\umax=1$, $2$ and $3$, and the differences decrease as $N^{-1/2}$. The two expansions therefore agree in every term through the constant, and Proposition~\ref{prop:F} is the specialisation of their result to the pseudosphere; conversely, the pseudosphere gives their axial potential and their constants a geometric meaning, the former being the curvature and the latter the profile of the boundary layers.

\subsection{Reading of the coefficients}\label{sec:reading}

At fixed density the expansion of Proposition~\ref{prop:F} has the structure (\ref{eq:Fstructure}). With $L_{b}=2\pi a(1+1/W)$ the total proper length (\ref{eq:lengths}) of the two horocycles and $n_{b}=N/(2\pi a^{2}\alpha)$, the term $B\sqrt N$ is
\begin{equation}
B\sqrt N=\frac{J}{\sqrt{2\pi}}\,L_{b}\sqrt{n_{b}}.\label{eq:tension}
\end{equation}
Substituting $\ln N=\ln n_{b}+\ln a^{2}+\ln2\pi\alpha$ in Eq.~(\ref{eq:expansion}), the terms $\tfrac12N\ln a^{2}$ and $-N\ln a^{2}$ of $\beta F$ combine with the term $N\ln(a/L_{0})$ of Eq.~(\ref{eq:FL0}) into $-N\ln L_{0}$. Measuring the areas of $Z_{N}$ in units of $L_{0}^{2}$, which adds $N\ln L_{0}^{2}$, and using Eq.~(\ref{eq:A2}),
\begin{align}
\beta F_{L_{0}}&=N\beta f+\beta\gamma\,L_{b}+D+O(N^{-1/2}),\label{eq:fixeddensity}\\
\beta f&=\frac12\ln\frac{n_{b}L_{0}^{2}}{2\pi^{2}}+1-\frac{\ln W}{W-1},\qquad
\beta\gamma=\frac{J}{\sqrt{2\pi}}\sqrt{n_{b}},\nonumber
\end{align}
The free energy is extensive, with a bulk term, a boundary term and a constant, and no logarithm. In $\beta f$ the first term is the free energy per particle of the flat plasma at $\Gamma=2$ \cite{Jancovici1981,AlastueyJancovici1981}, whose density dependence $\tfrac12\ln n_{b}$ is the origin of the term $\tfrac12N\ln N$ of Eq.~(\ref{eq:expansion}). The second is the average of $\ln w$ over the surface,
\begin{equation}
\frac1A\int_{S}\ln w\,\dd A=\frac1\alpha\int_{1}^{W}\frac{\ln w}{w^{2}}\,\dd w=1-\frac{\ln W}{W-1},\label{eq:avlnw}
\end{equation}
and it comes from the choice of the Coulomb potential: by the short-distance form of Sec.~\ref{sec:surface}, near a point of coordinate $w$ the potential $G-\ln(a/L_{0})$ is $-\ln(d/L_{0})-\ln w$, the planar potential with the constant $-\ln w$ added, and by the argument that led to Eq.~(\ref{eq:FL0}) a constant $c$ added to the potential changes $\beta F$ by $-c$ per particle. The boundary term is a surface tension $\beta\gamma=0.13471\sqrt{n_{b}}$ per unit length, the same at the rim and at the cut: the curvature enters only through the proper lengths, and locally each horocycle is a hard wall. This tension is that of the flat plasma against a hard wall. On the disc of the plane the norms are lower incomplete gamma functions $\gamma(k+1,N)$, and the same layer analysis gives the edge sum $S(N)=-\sum_{k<N}\ln[\gamma(k+1,N)/k!]=J\sqrt{2N}+O(1)$, which is Eq.~(\ref{eq:tension}) with $L_{b}=2\pi R$ and $n_{b}=N/\pi R^{2}$. Numerically, $[S(4N)-S(N)]/(\sqrt{8N}-\sqrt{2N})$ equals $0.337836$, $0.337710$ and $0.337679$ at $N=100$, $400$ and $1600$, converging to $J$. The constant $D$ collects a bulk part, $-\tfrac{13}{24}\alpha+(W-1)/12$, and the two layer constants; it depends on the shape and is not universal.

\section{The universal term}\label{sec:anomaly}

The absence of $\ln N$ is what the surface was chosen to test. A two-dimensional Coulomb system in its conducting phase screens charges in the bulk, but the charge fluctuations along a boundary decay only as the inverse square of the distance, and with respect to its finite-size corrections the system behaves as a critical one. Its long-distance sector is described by a conformal field theory of central charge $c=-1$ \cite{JMP1994,JancoviciTellez1996}. For a conformal field theory on a surface of linear size $L$, the trace anomaly produces a universal logarithmic term that couples to the geometry through the two curvature integrals \cite{CardyPeschel1988},
\begin{equation}
\beta F\supset-\frac{c}{12\pi}\Bigl[\int_{S}K\,\dd A+\oint_{\partial S}\kappa_{g}\,\dd s\Bigr]\ln L.\label{eq:anomaly}
\end{equation}
By the Gauss--Bonnet theorem (\ref{eq:GBgeneral}) the bracket is $2\pi\chi$, and with $c=-1$ and $\ln L=\tfrac12\ln N$ up to a constant, since $L\propto a\propto\sqrt N$ at fixed density,
\begin{equation}
\beta F\supset-\frac{c\chi}{6}\ln L=\frac{\chi}{12}\ln N.\label{eq:chi12}
\end{equation}
This is the term $\tfrac1{12}\ln N$ of the disc and $\tfrac16\ln N$ of the sphere \cite{JMP1994,JancoviciTellez1996,TellezForrester1999,SalazarTellez2016}. On the truncated pseudosphere the curvature integral is $-2\pi\alpha$ and the boundary integral $+2\pi\alpha$, Eq.~(\ref{eq:GB}); their sum vanishes, and Eq.~(\ref{eq:anomaly}) predicts no logarithm although each of its sources is present. If the two integrals entered Eq.~(\ref{eq:anomaly}) with different weights, the boundary integral being multiplied by $\lambda$, the bracket would be $2\pi\alpha(\lambda-1)$ and the coefficient of $\ln N$ would be $(\lambda-1)\alpha/12$. Counting the boundary twice ($\lambda=2$) gives $\alpha/12=0.029$, $0.061$ and $0.075$ for $\umax=1$, $2$ and $3$, and omitting it ($\lambda=0$) gives $-\alpha/12$. The estimates $\hat C$ of Table~\ref{tab:residual} are smaller in magnitude by about three orders of magnitude.

The derivation of Proposition~\ref{prop:F} shows how the cancellation happens. The curvature, present through the $1/w^{2}$ background and hence through the weights $w^{2\nu-2}$, produces the ratio of gamma functions (\ref{eq:sumlnc}), whose expansion (\ref{eq:lngammahalf}) has no logarithm; the two boundaries produce layers that contribute $\sqrt N$ and constants. The curvature does not disappear from the problem: it appears in $A_{2}$, in the layer constants and in the finite-$N$ density. Its contribution to the universal term is cancelled by the boundary geometry in exactly the combination fixed by the Gauss--Bonnet theorem.

\section{Contact densities}\label{sec:contact}

\begin{proposition}\label{prop:contact}
At both horocycles the density tends to the contact value of the flat plasma against a hard wall,
\begin{equation}
\lim_{N\to\infty}\frac{n(1)}{n_{b}}=\lim_{N\to\infty}\frac{n(W)}{n_{b}}=\ln2.\label{eq:contact}
\end{equation}
\end{proposition}

\begin{proof}
With the gamma density $f_{s}(x)=x^{s-1}e^{-x}/\Gamma(s)$ and the bracket (\ref{eq:bk}), Eqs.~(\ref{eq:density}) and (\ref{eq:lnIk}) give $n(1)/n_{b}=\nu^{-1}\sum_{k}c_{k}f_{s}(c_{k})/b_{k}$. Only the rim layer contributes, $f_{s}(c_{k})$ being exponentially small for the other orbitals. In the layer, the gamma density is a Gaussian of variance $s$ about its peak, and $(c_{k}-s)^{2}/(2s)=2k^{2}/s=t_{k}^{2}$, so that $f_{s}(c_{k})=(2\pi s)^{-1/2}e^{-t_{k}^{2}}[1+O(s^{-1/2})]$; moreover $c_{k}=s[1+O(s^{-1/2})]$ and $b_{k}=u(t_{k})+O(s^{-1/2})$. Therefore
\begin{equation}
\frac{n(1)}{n_{b}}=\frac{s}{\nu}\,\frac{1}{\sqrt{2\pi s}}\sum_{k\ge0}\frac{e^{-t_{k}^{2}}}{u(t_{k})}\bigl[1+O(s^{-1/2})\bigr].\nonumber
\end{equation}
Equation~(\ref{eq:EM}) with the step $\sqrt{2/s}$ turns the sum into $\sqrt{s/2}\int_{0}^{\infty}e^{-t^{2}}u^{-1}\dd t+O(1)$, and with $s/\nu\to2$ the prefactor and the step combine into $2(2\pi s)^{-1/2}\sqrt{s/2}=\pi^{-1/2}$:
\begin{equation}
\frac{n(1)}{n_{b}}=\frac{1}{\sqrt\pi}\int_{0}^{\infty}\frac{e^{-t^{2}}}{u(t)}\,\dd t\,\bigl[1+O(s^{-1/2})\bigr].\label{eq:contactint}
\end{equation}
Since $\dd u=\pi^{-1/2}e^{-t^{2}}\dd t$ and $u$ runs from $\tfrac12$ to $1$, the integral equals $\sqrt\pi\ln2$. At the cut, $W^{2\nu}=W^{s+1}$ gives $n(W)/n_{b}=\nu^{-1}\sum_{k}W(c_{k}W)f_{s}(c_{k}W)/b_{k}$, and the same computation with step $W\sqrt{2/s}$ gives the same limit: the factor $W$ cancels against the coarser step.
\end{proof}

\begin{figure}[t]
\centering
\includegraphics[width=\columnwidth]{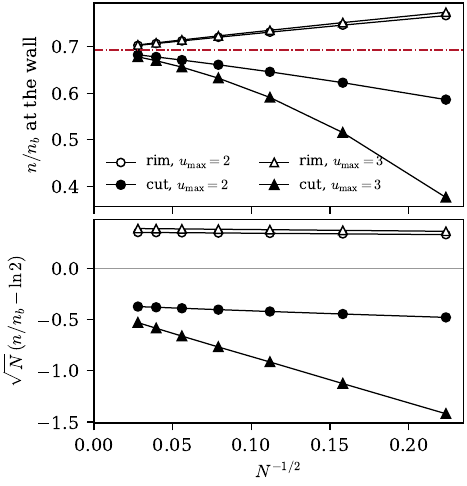}
\caption{Contact densities from the exact density (\ref{eq:density}) for $N=20,\dots,1280$. Top: rim (open symbols) and cut (filled symbols) against $N^{-1/2}$ for $\umax=2$ and $3$; the dash-dotted line is $\ln2$. Bottom: the scaled deviations $\sqrt N\,(n/n_{b}-\ln2)$, which tend to constants of opposite sign at the two walls.}
\label{fig:contact}
\end{figure}

The value $n_{b}\ln2$ is the contact density of the flat plasma at $\Gamma=2$ against a hard wall \cite{Jancovici1982}, consistent with the contact theorem relating the contact density to the bulk pressure \cite{CFG1980}. Figure~\ref{fig:contact} shows that the two walls approach it from opposite sides, as $N^{-1/2}$. At the rim the first orbital sits exactly at the wall ($t_{0}=0$), while at the cut it is offset into the interior by $t_{0}=(W+1)/\sqrt{2s}$, so that the cut is the emptier wall at every finite $N$. At $\umax=2$ the scaled deviation $\sqrt N\,(n/n_{b}-\ln2)$ is $+0.351$, $+0.352$ and $+0.353$ at the rim for $N=640$, $1280$ and $2560$, and $-0.379$, $-0.372$ and $-0.367$ at the cut, the latter still drifting. These $N^{-1/2}$ coefficients are not derived in the present contribution.

\section{Beyond $\Gamma=2$: Monte Carlo simulation}\label{sec:gamma}

The exact solution is specific to $\Gamma=2$, and it leaves open which of its features follow from the determinantal structure and which are properties of the plasma on this surface at any coupling. The conformal reduction of Sec.~\ref{sec:surface} holds at every $\Gamma$: the plasma is a system on the flat cylinder $[1,W]\times S^{1}$ with the pair energy $q^{2}G$ of Eq.~(\ref{eq:G}), the one-body energy $qv_{b}$ of Eq.~(\ref{eq:vb}), hard walls at $w=1$ and $w=W$, and the measure $(a^{2}/w^{2})\,\dd w\,\dd\phi$. This form suits a Metropolis simulation \cite{Metropolis1953,CLWH1982}. A particle is moved in $(w,\phi)$, the energy change $\Delta E$ is computed with Eq.~(\ref{eq:Greal}), and the move is accepted with probability $\min\{1,e^{-\Gamma\Delta E/q^{2}}(w_{\mathrm{old}}/w_{\mathrm{new}})^{2}\}$, the last factor being the ratio of area elements; moves outside $[1,W]$ are rejected. The density is accumulated in bins of $u$ and normalized by the background. Figure~\ref{fig:system}(a,b) shows an equilibrium configuration produced in this way.

\begin{figure}[t]
\centering
\includegraphics[width=\columnwidth]{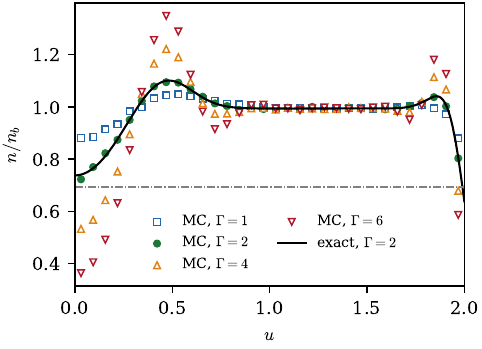}
\caption{Density profiles from Metropolis simulation at $\umax=2$ and $N=60$ for $\Gamma=1$, $2$, $4$, $6$ ($2\times10^{5}$ sweeps after $2\times10^{4}$ of equilibration, 32 bins in $u$), and the exact density (\ref{eq:density}) at $\Gamma=2$ (solid line). The dash-dotted line is $\ln2$. At $\Gamma=2$ the simulation reproduces the exact profile with a root-mean-square deviation of $0.005\,n_{b}$.}
\label{fig:mc}
\end{figure}

\begin{figure}[t]
\centering
\includegraphics[width=\columnwidth]{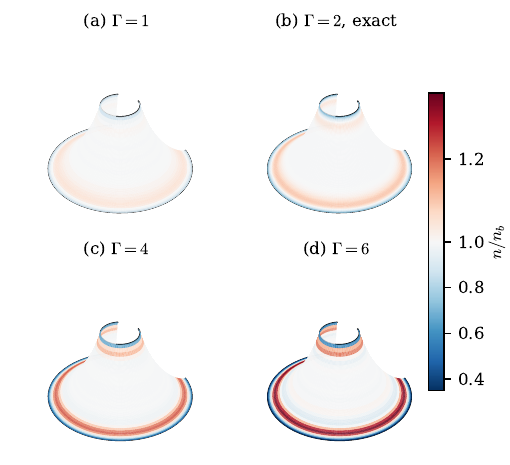}
\caption{Densities at $\umax=2$ and $N=60$ drawn on the surface, with a common colour scale centred on the bulk value: Monte Carlo profiles of Fig.~\ref{fig:mc} for (a) $\Gamma=1$, (c) $\Gamma=4$ and (d) $\Gamma=6$, in which each of the 32 bins in $u$ is a band of uniform colour, and (b) the exact density (\ref{eq:density}) at $\Gamma=2$. The rim is the wide circle at the bottom and the cut the narrow one at the top.}
\label{fig:mc3d}
\end{figure}

Figure~\ref{fig:mc} shows the profiles at $\umax=2$ and $N=60$. At $\Gamma=2$ the simulation reproduces the exact density: the root-mean-square deviation over the 32 bins is $0.005\,n_{b}$ and the largest deviation, in a bin adjacent to a wall, is below $0.02\,n_{b}$, consistent with the statistical error. This validates the code before the other couplings are read. At $\Gamma=1$ the depletion is shallower and the overshoot weaker. At $\Gamma=4$ and $6$ the wall layers deepen, the overshoot becomes a pronounced first peak, and oscillations with the period of the interparticle spacing extend into the interior, an oscillatory layering that develops with increasing coupling. In all the simulated cases, $\Gamma=1$, $2$, $4$ and $6$ at $N=60$ and $\umax=2$, the two walls remain unequal. The simulation at $\Gamma=2$ serves only to validate the code against the exact solution; at the other couplings no exact result is available, and the simulation is the only source of the profiles. Figure~\ref{fig:mc3d} shows them on the surface, with the exact density in place of the simulation at $\Gamma=2$: as the coupling grows, the depleted strip at each wall deepens, the overshoot sharpens into a ring, and at $\Gamma=6$ a second, weaker depleted ring follows it.

The bins are uniform in $u$, and their proper widths are not. The first bin, at the rim, has the meridional width $a\ln\cosh(\umax/32)\simeq0.002\,a$, far below the spacing $n_{b}^{-1/2}\simeq0.28\,a$, so that its content is the contact density. The last bin, at the cut, has the width $a\ln[\cosh\umax/\cosh(31\umax/32)]\simeq0.06\,a$ and averages the steep cut layer over a fifth of a spacing. The two end bins are therefore compared across couplings, not with each other. Because the first bin is so narrow it collects few counts, and its statistical uncertainty is the largest of the profile: two independent implementations of the algorithm, one in Python with a compiled inner loop and one in C, run with different random sequences, give $0.722$ and $0.755\,n_{b}$ in that bin at $\Gamma=2$, on either side of the exact value $0.740$ at its centre, so that the end-bin values carry an uncertainty of about $0.02\,n_{b}$. Averaged over the two runs, the rim contact density falls monotonically with the coupling, $0.88$, $0.74$, $0.53$ and $0.36\,n_{b}$ at $\Gamma=1$, $2$, $4$ and $6$, the exact contact value at $\Gamma=2$ being $0.737$. The average over the last bin falls more slowly, $0.88$, $0.80$, $0.68$ and $0.58\,n_{b}$; the exact value at $\Gamma=2$ at the centre of that bin is $0.806$. The heights of the first peak, $1.22$ and $1.35\,n_{b}$ at $\Gamma=4$ and $6$, agree between the two runs to $0.004\,n_{b}$, and their mean potential energies per particle agree within $5\times10^{-4}$ in units of $q^{2}$ at every coupling. Three questions answered exactly at $\Gamma=2$ remain open at other couplings and can be addressed with the same method: whether the two contact densities tend to a common $\Gamma$-dependent limit, as the contact theorem \cite{CFG1980} suggests through the bulk pressure; whether the boundary free energy remains the flat hard-wall tension times the total proper length, which thermodynamic integration in $\Gamma$ can test; and whether the absence of a $\ln N$ term holds throughout the fluid phase, as Eq.~(\ref{eq:anomaly}) predicts. At $\Gamma=4$ and $6$ the expansions of the powers of the Vandermonde determinant \cite{SamajPercus1995,TellezForrester1999,SalazarTellez2016} offer an independent exact check at small $N$.

\section{Conclusions}\label{sec:conclusions}

The two-dimensional one-component plasma has been solved exactly at the coupling $\Gamma=2$ on the truncated Beltrami pseudosphere, starting from the Hamiltonian of the charges on the curved surface. The kinetic energy contributes only the thermal wavelength, and the curvature enters the configurational problem through the area element. The coordinate $w=\cosh u$ maps the surface onto a flat cylinder with conformal factor $a^{2}/w^{2}$; because the two-dimensional Coulomb potential is unchanged by such a map, the particles interact as on a flat cylinder, and the negative curvature enters the plasma through the area element and, with it, through the background, uniform on the surface and therefore proportional to $1/w^{2}$ on the cylinder. At $\Gamma=2$ the Boltzmann factor is the squared modulus of a Vandermonde determinant in the variables $e^{\zeta_{i}}$, the mechanism that also underlies the free-fermion and Landau-level pictures of the plasma. Neutrality removes the position of the cut from the one-body weights, and the partition function becomes a product of incomplete gamma functions, incomplete because the surface has two boundaries. Each orbital occupies a band of the surface, and only the orbitals cut by a wall feel the boundaries.

Because the area of a surface of given curvature is bounded, the thermodynamic limit has been taken at fixed density on surfaces of the same shape and growing size, as on the sphere; since the logarithmic interaction has no intrinsic length, this is equivalent to letting $N$ grow on a surface of fixed size, the form in which the calculation is done. The free energy has been obtained asymptotically through the constant term, with every coefficient in closed form; the remainder is consistent with $O(N^{-1/2})$ numerically but is not bounded rigorously. At fixed density the free energy is extensive: the bulk term is the free energy of the flat plasma at $\Gamma=2$ plus the average of $\ln w$ over the surface, which comes from the choice of the Coulomb potential; the coefficient of $\sqrt N$ is the surface tension of the flat plasma against a hard wall, given by a single quadrature, times the total proper length of the two boundaries; and the constant follows from the second term of Temme's uniform expansion. There is no logarithmic term. On a surface whose integrated curvature and integrated boundary curvature are both of order one, this is the content of the conformal-anomaly prediction with Euler characteristic zero, and the exact solution shows its mechanism: the curvature produces a ratio of gamma functions whose expansion has no logarithm, and the boundaries produce layers that contribute only $\sqrt N$ and constants. Every coefficient has been checked against the exact free energy up to $N=4096$; after removal of the leading correction the constant agrees to four or five digits, and the estimated coefficient of a possible logarithmic term is of order $10^{-5}$ at the largest sizes, about three orders of magnitude below the coefficient $\alpha/12$ that a doubled boundary term would give. In the conformal coordinates the system is a cylinder plasma with the axial potential $A(\ln r-\ln\ln r)$ and walls at the edges of its droplet, and the expansion agrees term by term with the general cylinder result of Shen and Forrester \cite{ShenForrester2026}, of which it is a special case; the geometric content, the curvature as the potential and the Gauss--Bonnet cancellation as the reason for the vanishing logarithm, is what the pseudosphere adds.

The curvature remains visible. It appears in the finite-$N$ density, where the two walls, locally identical horocycles, are depleted differently because their circumferences differ by the factor $W$; in the coefficient $A_{2}$ and in the constant $D$; and in the proper lengths that multiply the surface tension. Both contact densities tend to the flat value $n_{b}\ln2$, from opposite sides, with corrections of order $N^{-1/2}$. Metropolis simulations, validated against the exact density at $\Gamma=2$, show that the boundary layers deepen and the first density peak grows with the coupling, and that the two walls remain unequal at $\Gamma=1$, $4$ and $6$ for the size and shape simulated. Whether the common contact limit, the flat surface tension and the absence of a logarithmic term persist throughout the fluid phase is open, and the simulation method of Sec.~\ref{sec:gamma} can address it. The two-component plasma on the same surface is not treated: unlike the one-component plasma, it requires an external potential to be solvable \cite{FantoniReview}. The truncated pseudosphere is a physical object, fabricated as a microlaser and proposed as a graphene substrate, and the exact solution presented here is a reference against which measurements or simulations of charges on that surface can be read.

\begin{acknowledgments}
J.B.\ appreciates fruitful discussions with Maria Vallespir-Socias, Joana Rossell\'o, Margalida Batle, Maria del Mar Batle, and Regina Batle. J.B.\ received no funding for the present work.
\end{acknowledgments}

\section*{Note added}
While this work was being completed we became aware of the work of Shen and Forrester, who obtained the large-$N$ expansion of the free energy of the plasma at $\Gamma=2$ on the surface of a cylinder, with an arbitrary one-body potential along the axis and hard walls at the edges of the droplet or inside it, and showed that it contains no term in $\ln N$ in any of these cases \cite{ShenForrester2026}. In the conformal coordinates of Sec.~\ref{sec:surface}, the plasma on the truncated pseudosphere is a cylinder plasma with a particular axial potential and with its two walls at the edges of the droplet, so that the absence of the logarithm found here is a special case of their result. The two expansions are compared term by term in Sec.~\ref{sec:largeN}, where they are found to agree, and where two of their universal constants are identified with the quadratures of the boundary layers of the present problem. What the pseudosphere adds is the geometric reading: the axial potential is the curvature, and the vanishing of the logarithm is the cancellation of the two curvature integrals in the Gauss--Bonnet theorem.

\section*{Data Availability}
The data that support the findings of this article are available from the author on reasonable request. They include the computer-algebra worksheet that verifies the identities of Secs.~\ref{sec:surface}--\ref{sec:largeN}, the code that evaluates the exact solution and produces Figs.~\ref{fig:orbitals}--\ref{fig:contact} and Table~\ref{tab:residual}, and the two Monte Carlo implementations of Sec.~\ref{sec:gamma}.

\bibliography{ocp_refs}

\begin{thebibliography}{38}%
\makeatletter
\providecommand \@ifxundefined [1]{%
 \@ifx{#1\undefined}
}%
\providecommand \@ifnum [1]{%
 \ifnum #1\expandafter \@firstoftwo
 \else \expandafter \@secondoftwo
 \fi
}%
\providecommand \@ifx [1]{%
 \ifx #1\expandafter \@firstoftwo
 \else \expandafter \@secondoftwo
 \fi
}%
\providecommand \natexlab [1]{#1}%
\providecommand \enquote  [1]{``#1''}%
\providecommand \bibnamefont  [1]{#1}%
\providecommand \bibfnamefont [1]{#1}%
\providecommand \citenamefont [1]{#1}%
\providecommand \href@noop [0]{\@secondoftwo}%
\providecommand \href [0]{\begingroup \@sanitize@url \@href}%
\providecommand \@href[1]{\@@startlink{#1}\@@href}%
\providecommand \@@href[1]{\endgroup#1\@@endlink}%
\providecommand \@sanitize@url [0]{\catcode `\\12\catcode `\$12\catcode
  `\&12\catcode `\#12\catcode `\^12\catcode `\_12\catcode `\%12\relax}%
\providecommand \@@startlink[1]{}%
\providecommand \@@endlink[0]{}%
\providecommand \url  [0]{\begingroup\@sanitize@url \@url }%
\providecommand \@url [1]{\endgroup\@href {#1}{\urlprefix }}%
\providecommand \urlprefix  [0]{URL }%
\providecommand \Eprint [0]{\href }%
\providecommand \doibase [0]{https://doi.org/}%
\providecommand \selectlanguage [0]{\@gobble}%
\providecommand \bibinfo  [0]{\@secondoftwo}%
\providecommand \bibfield  [0]{\@secondoftwo}%
\providecommand \translation [1]{[#1]}%
\providecommand \BibitemOpen [0]{}%
\providecommand \bibitemStop [0]{}%
\providecommand \bibitemNoStop [0]{.\EOS\space}%
\providecommand \EOS [0]{\spacefactor3000\relax}%
\providecommand \BibitemShut  [1]{\csname bibitem#1\endcsname}%
\let\auto@bib@innerbib\@empty
\bibitem [{\citenamefont {Jancovici}(1995)}]{Jancovici1995}%
  \BibitemOpen
  \bibfield  {author} {\bibinfo {author} {\bibfnamefont {B.}~\bibnamefont
  {Jancovici}},\ }\bibfield  {title} {\bibinfo {title} {Classical {Coulomb}
  systems: screening and correlations revisited},\ }\href@noop {} {\bibfield
  {journal} {\bibinfo  {journal} {J. Stat. Phys.}\ }\textbf {\bibinfo {volume}
  {80}},\ \bibinfo {pages} {445} (\bibinfo {year} {1995})}\BibitemShut
  {NoStop}%
\bibitem [{\citenamefont {Forrester}(1998)}]{Forrester1998}%
  \BibitemOpen
  \bibfield  {author} {\bibinfo {author} {\bibfnamefont {P.~J.}\ \bibnamefont
  {Forrester}},\ }\bibfield  {title} {\bibinfo {title} {Exact results for
  two-dimensional {Coulomb} systems},\ }\href@noop {} {\bibfield  {journal}
  {\bibinfo  {journal} {Phys. Rep.}\ }\textbf {\bibinfo {volume} {301}},\
  \bibinfo {pages} {235} (\bibinfo {year} {1998})}\BibitemShut {NoStop}%
\bibitem [{\citenamefont {Forrester}(2010)}]{ForresterBook}%
  \BibitemOpen
  \bibfield  {author} {\bibinfo {author} {\bibfnamefont {P.~J.}\ \bibnamefont
  {Forrester}},\ }\href@noop {} {\emph {\bibinfo {title} {Log-Gases and Random
  Matrices}}}\ (\bibinfo  {publisher} {Princeton University Press},\ \bibinfo
  {address} {Princeton},\ \bibinfo {year} {2010})\BibitemShut {NoStop}%
\bibitem [{\citenamefont {Caillol}\ \emph {et~al.}(1982)\citenamefont
  {Caillol}, \citenamefont {Levesque}, \citenamefont {Weis},\ and\
  \citenamefont {Hansen}}]{CLWH1982}%
  \BibitemOpen
  \bibfield  {author} {\bibinfo {author} {\bibfnamefont {J.~M.}\ \bibnamefont
  {Caillol}}, \bibinfo {author} {\bibfnamefont {D.}~\bibnamefont {Levesque}},
  \bibinfo {author} {\bibfnamefont {J.~J.}\ \bibnamefont {Weis}},\ and\
  \bibinfo {author} {\bibfnamefont {J.~P.}\ \bibnamefont {Hansen}},\ }\bibfield
   {title} {\bibinfo {title} {A {Monte Carlo} study of the classical
  two-dimensional one-component plasma},\ }\href@noop {} {\bibfield  {journal}
  {\bibinfo  {journal} {J. Stat. Phys.}\ }\textbf {\bibinfo {volume} {28}},\
  \bibinfo {pages} {325} (\bibinfo {year} {1982})}\BibitemShut {NoStop}%
\bibitem [{\citenamefont {Jancovici}(1981)}]{Jancovici1981}%
  \BibitemOpen
  \bibfield  {author} {\bibinfo {author} {\bibfnamefont {B.}~\bibnamefont
  {Jancovici}},\ }\bibfield  {title} {\bibinfo {title} {Exact results for the
  two-dimensional one-component plasma},\ }\href@noop {} {\bibfield  {journal}
  {\bibinfo  {journal} {Phys. Rev. Lett.}\ }\textbf {\bibinfo {volume} {46}},\
  \bibinfo {pages} {386} (\bibinfo {year} {1981})}\BibitemShut {NoStop}%
\bibitem [{\citenamefont {Alastuey}\ and\ \citenamefont
  {Jancovici}(1981)}]{AlastueyJancovici1981}%
  \BibitemOpen
  \bibfield  {author} {\bibinfo {author} {\bibfnamefont {A.}~\bibnamefont
  {Alastuey}}\ and\ \bibinfo {author} {\bibfnamefont {B.}~\bibnamefont
  {Jancovici}},\ }\bibfield  {title} {\bibinfo {title} {On the classical
  two-dimensional one-component {Coulomb} plasma},\ }\href@noop {} {\bibfield
  {journal} {\bibinfo  {journal} {J. Phys. (France)}\ }\textbf {\bibinfo
  {volume} {42}},\ \bibinfo {pages} {1} (\bibinfo {year} {1981})}\BibitemShut
  {NoStop}%
\bibitem [{\citenamefont {Ginibre}(1965)}]{Ginibre1965}%
  \BibitemOpen
  \bibfield  {author} {\bibinfo {author} {\bibfnamefont {J.}~\bibnamefont
  {Ginibre}},\ }\bibfield  {title} {\bibinfo {title} {Statistical ensembles of
  complex, quaternion, and real matrices},\ }\href@noop {} {\bibfield
  {journal} {\bibinfo  {journal} {J. Math. Phys.}\ }\textbf {\bibinfo {volume}
  {6}},\ \bibinfo {pages} {440} (\bibinfo {year} {1965})}\BibitemShut {NoStop}%
\bibitem [{\citenamefont {Laughlin}(1983)}]{Laughlin1983}%
  \BibitemOpen
  \bibfield  {author} {\bibinfo {author} {\bibfnamefont {R.~B.}\ \bibnamefont
  {Laughlin}},\ }\bibfield  {title} {\bibinfo {title} {Anomalous quantum {Hall}
  effect: An incompressible quantum fluid with fractionally charged
  excitations},\ }\href@noop {} {\bibfield  {journal} {\bibinfo  {journal}
  {Phys. Rev. Lett.}\ }\textbf {\bibinfo {volume} {50}},\ \bibinfo {pages}
  {1395} (\bibinfo {year} {1983})}\BibitemShut {NoStop}%
\bibitem [{\citenamefont {Choquard}\ \emph {et~al.}(1983)\citenamefont
  {Choquard}, \citenamefont {Forrester},\ and\ \citenamefont
  {Smith}}]{CFS1983}%
  \BibitemOpen
  \bibfield  {author} {\bibinfo {author} {\bibfnamefont {P.}~\bibnamefont
  {Choquard}}, \bibinfo {author} {\bibfnamefont {P.~J.}\ \bibnamefont
  {Forrester}},\ and\ \bibinfo {author} {\bibfnamefont {E.~R.}\ \bibnamefont
  {Smith}},\ }\bibfield  {title} {\bibinfo {title} {The two-dimensional
  one-component plasma at {$\Gamma=2$}: The semiperiodic strip},\ }\href@noop
  {} {\bibfield  {journal} {\bibinfo  {journal} {J. Stat. Phys.}\ }\textbf
  {\bibinfo {volume} {33}},\ \bibinfo {pages} {13} (\bibinfo {year}
  {1983})}\BibitemShut {NoStop}%
\bibitem [{\citenamefont {Caillol}(1981)}]{Caillol1981}%
  \BibitemOpen
  \bibfield  {author} {\bibinfo {author} {\bibfnamefont {J.~M.}\ \bibnamefont
  {Caillol}},\ }\bibfield  {title} {\bibinfo {title} {Exact results for a
  two-dimensional one-component plasma on a sphere},\ }\href@noop {} {\bibfield
   {journal} {\bibinfo  {journal} {J. Phys. Lett. (France)}\ }\textbf {\bibinfo
  {volume} {42}},\ \bibinfo {pages} {L245} (\bibinfo {year}
  {1981})}\BibitemShut {NoStop}%
\bibitem [{\citenamefont {Jancovici}\ and\ \citenamefont
  {T{\'e}llez}(1998)}]{JancoviciTellez1998}%
  \BibitemOpen
  \bibfield  {author} {\bibinfo {author} {\bibfnamefont {B.}~\bibnamefont
  {Jancovici}}\ and\ \bibinfo {author} {\bibfnamefont {G.}~\bibnamefont
  {T{\'e}llez}},\ }\bibfield  {title} {\bibinfo {title} {Two-dimensional
  {Coulomb} systems on a surface of constant negative curvature},\ }\href@noop
  {} {\bibfield  {journal} {\bibinfo  {journal} {J. Stat. Phys.}\ }\textbf
  {\bibinfo {volume} {91}},\ \bibinfo {pages} {953} (\bibinfo {year}
  {1998})}\BibitemShut {NoStop}%
\bibitem [{\citenamefont {Fantoni}\ \emph {et~al.}(2003)\citenamefont
  {Fantoni}, \citenamefont {Jancovici},\ and\ \citenamefont
  {T{\'e}llez}}]{FJT2003}%
  \BibitemOpen
  \bibfield  {author} {\bibinfo {author} {\bibfnamefont {R.}~\bibnamefont
  {Fantoni}}, \bibinfo {author} {\bibfnamefont {B.}~\bibnamefont {Jancovici}},\
  and\ \bibinfo {author} {\bibfnamefont {G.}~\bibnamefont {T{\'e}llez}},\
  }\bibfield  {title} {\bibinfo {title} {Pressures for a one-component plasma
  on a pseudosphere},\ }\href@noop {} {\bibfield  {journal} {\bibinfo
  {journal} {J. Stat. Phys.}\ }\textbf {\bibinfo {volume} {112}},\ \bibinfo
  {pages} {27} (\bibinfo {year} {2003})}\BibitemShut {NoStop}%
\bibitem [{\citenamefont {Fantoni}\ and\ \citenamefont
  {T{\'e}llez}(2008)}]{FantoniTellez2008}%
  \BibitemOpen
  \bibfield  {author} {\bibinfo {author} {\bibfnamefont {R.}~\bibnamefont
  {Fantoni}}\ and\ \bibinfo {author} {\bibfnamefont {G.}~\bibnamefont
  {T{\'e}llez}},\ }\bibfield  {title} {\bibinfo {title} {Two-dimensional
  one-component plasma on {Flamm's} paraboloid},\ }\href@noop {} {\bibfield
  {journal} {\bibinfo  {journal} {J. Stat. Phys.}\ }\textbf {\bibinfo {volume}
  {133}},\ \bibinfo {pages} {449} (\bibinfo {year} {2008})}\BibitemShut
  {NoStop}%
\bibitem [{\citenamefont {Fantoni}(2019)}]{FantoniReview}%
  \BibitemOpen
  \bibfield  {author} {\bibinfo {author} {\bibfnamefont {R.}~\bibnamefont
  {Fantoni}},\ }\href@noop {} {\bibinfo {title} {Plasma living in a curved
  surface at some special temperature}} (\bibinfo {year} {2019}),\ \Eprint
  {https://arxiv.org/abs/1904.04066} {arXiv:1904.04066} \BibitemShut {NoStop}%
\bibitem [{\citenamefont {Jancovici}\ \emph {et~al.}(1994)\citenamefont
  {Jancovici}, \citenamefont {Manificat},\ and\ \citenamefont
  {Pisani}}]{JMP1994}%
  \BibitemOpen
  \bibfield  {author} {\bibinfo {author} {\bibfnamefont {B.}~\bibnamefont
  {Jancovici}}, \bibinfo {author} {\bibfnamefont {G.}~\bibnamefont
  {Manificat}},\ and\ \bibinfo {author} {\bibfnamefont {C.}~\bibnamefont
  {Pisani}},\ }\bibfield  {title} {\bibinfo {title} {{Coulomb} systems seen as
  critical systems: Finite-size effects in two dimensions},\ }\href@noop {}
  {\bibfield  {journal} {\bibinfo  {journal} {J. Stat. Phys.}\ }\textbf
  {\bibinfo {volume} {76}},\ \bibinfo {pages} {307} (\bibinfo {year}
  {1994})}\BibitemShut {NoStop}%
\bibitem [{\citenamefont {Jancovici}\ and\ \citenamefont
  {T{\'e}llez}(1996)}]{JancoviciTellez1996}%
  \BibitemOpen
  \bibfield  {author} {\bibinfo {author} {\bibfnamefont {B.}~\bibnamefont
  {Jancovici}}\ and\ \bibinfo {author} {\bibfnamefont {G.}~\bibnamefont
  {T{\'e}llez}},\ }\bibfield  {title} {\bibinfo {title} {{Coulomb} systems seen
  as critical systems: Ideal conductor boundaries},\ }\href@noop {} {\bibfield
  {journal} {\bibinfo  {journal} {J. Stat. Phys.}\ }\textbf {\bibinfo {volume}
  {82}},\ \bibinfo {pages} {609} (\bibinfo {year} {1996})}\BibitemShut
  {NoStop}%
\bibitem [{\citenamefont {Cardy}\ and\ \citenamefont
  {Peschel}(1988)}]{CardyPeschel1988}%
  \BibitemOpen
  \bibfield  {author} {\bibinfo {author} {\bibfnamefont {J.~L.}\ \bibnamefont
  {Cardy}}\ and\ \bibinfo {author} {\bibfnamefont {I.}~\bibnamefont
  {Peschel}},\ }\bibfield  {title} {\bibinfo {title} {Finite-size dependence of
  the free energy in two-dimensional critical systems},\ }\href@noop {}
  {\bibfield  {journal} {\bibinfo  {journal} {Nucl. Phys. B}\ }\textbf
  {\bibinfo {volume} {300}},\ \bibinfo {pages} {377} (\bibinfo {year}
  {1988})}\BibitemShut {NoStop}%
\bibitem [{\citenamefont {T{\'e}llez}\ and\ \citenamefont
  {Forrester}(1999)}]{TellezForrester1999}%
  \BibitemOpen
  \bibfield  {author} {\bibinfo {author} {\bibfnamefont {G.}~\bibnamefont
  {T{\'e}llez}}\ and\ \bibinfo {author} {\bibfnamefont {P.~J.}\ \bibnamefont
  {Forrester}},\ }\bibfield  {title} {\bibinfo {title} {Exact finite-size study
  of the {2D OCP} at {$\Gamma=4$} and {$\Gamma=6$}},\ }\href@noop {} {\bibfield
   {journal} {\bibinfo  {journal} {J. Stat. Phys.}\ }\textbf {\bibinfo {volume}
  {97}},\ \bibinfo {pages} {489} (\bibinfo {year} {1999})}\BibitemShut
  {NoStop}%
\bibitem [{\citenamefont {Salazar}\ and\ \citenamefont
  {T{\'e}llez}(2016)}]{SalazarTellez2016}%
  \BibitemOpen
  \bibfield  {author} {\bibinfo {author} {\bibfnamefont {R.}~\bibnamefont
  {Salazar}}\ and\ \bibinfo {author} {\bibfnamefont {G.}~\bibnamefont
  {T{\'e}llez}},\ }\bibfield  {title} {\bibinfo {title} {Exact energy
  computation of the one component plasma on a sphere for even values of the
  coupling parameter},\ }\href@noop {} {\bibfield  {journal} {\bibinfo
  {journal} {J. Stat. Phys.}\ }\textbf {\bibinfo {volume} {164}},\ \bibinfo
  {pages} {969} (\bibinfo {year} {2016})}\BibitemShut {NoStop}%
\bibitem [{\citenamefont {Zabrodin}\ and\ \citenamefont
  {Wiegmann}(2006)}]{ZabrodinWiegmann2006}%
  \BibitemOpen
  \bibfield  {author} {\bibinfo {author} {\bibfnamefont {A.}~\bibnamefont
  {Zabrodin}}\ and\ \bibinfo {author} {\bibfnamefont {P.}~\bibnamefont
  {Wiegmann}},\ }\bibfield  {title} {\bibinfo {title} {Large-{$N$} expansion
  for the {2D Dyson} gas},\ }\href@noop {} {\bibfield  {journal} {\bibinfo
  {journal} {J. Phys. A: Math. Gen.}\ }\textbf {\bibinfo {volume} {39}},\
  \bibinfo {pages} {8933} (\bibinfo {year} {2006})}\BibitemShut {NoStop}%
\bibitem [{\citenamefont {Klevtsov}(2014)}]{Klevtsov2014}%
  \BibitemOpen
  \bibfield  {author} {\bibinfo {author} {\bibfnamefont {S.}~\bibnamefont
  {Klevtsov}},\ }\bibfield  {title} {\bibinfo {title} {Random normal matrices,
  {Bergman} kernel and projective embeddings},\ }\href
  {https://doi.org/10.1007/JHEP01(2014)133} {\bibfield  {journal} {\bibinfo
  {journal} {J. High Energy Phys.}\ }\textbf {\bibinfo {volume} {2014}}\bibinfo
   {number} { (01)},\ \bibinfo {pages} {133}}\BibitemShut {NoStop}%
\bibitem [{\citenamefont {Can}\ \emph {et~al.}(2014)\citenamefont {Can},
  \citenamefont {Laskin},\ and\ \citenamefont
  {Wiegmann}}]{CanLaskinWiegmann2014}%
  \BibitemOpen
\bibfield  {number} {  }\bibfield  {author} {\bibinfo {author} {\bibfnamefont
  {T.}~\bibnamefont {Can}}, \bibinfo {author} {\bibfnamefont {M.}~\bibnamefont
  {Laskin}},\ and\ \bibinfo {author} {\bibfnamefont {P.}~\bibnamefont
  {Wiegmann}},\ }\bibfield  {title} {\bibinfo {title} {Fractional quantum
  {Hall} effect in a curved space: Gravitational anomaly and electromagnetic
  response},\ }\href@noop {} {\bibfield  {journal} {\bibinfo  {journal} {Phys.
  Rev. Lett.}\ }\textbf {\bibinfo {volume} {113}},\ \bibinfo {pages} {046803}
  (\bibinfo {year} {2014})}\BibitemShut {NoStop}%
\bibitem [{\citenamefont {Ferrari}\ and\ \citenamefont
  {Klevtsov}(2014)}]{FerrariKlevtsov2014}%
  \BibitemOpen
  \bibfield  {author} {\bibinfo {author} {\bibfnamefont {F.}~\bibnamefont
  {Ferrari}}\ and\ \bibinfo {author} {\bibfnamefont {S.}~\bibnamefont
  {Klevtsov}},\ }\bibfield  {title} {\bibinfo {title} {{FQHE} on curved
  backgrounds, free fields and large {$N$}},\ }\href
  {https://doi.org/10.1007/JHEP12(2014)086} {\bibfield  {journal} {\bibinfo
  {journal} {J. High Energy Phys.}\ }\textbf {\bibinfo {volume} {2014}}\bibinfo
   {number} { (12)},\ \bibinfo {pages} {086}}\BibitemShut {NoStop}%
\bibitem [{\citenamefont {Beltrami}(1868)}]{Beltrami1868}%
  \BibitemOpen
\bibfield  {number} {  }\bibfield  {author} {\bibinfo {author} {\bibfnamefont
  {E.}~\bibnamefont {Beltrami}},\ }\bibfield  {title} {\bibinfo {title} {Saggio
  di interpretazione della geometria non-euclidea},\ }\href@noop {} {\bibfield
  {journal} {\bibinfo  {journal} {Giorn. Mat.}\ }\textbf {\bibinfo {volume}
  {6}},\ \bibinfo {pages} {284} (\bibinfo {year} {1868})}\BibitemShut {NoStop}%
\bibitem [{\citenamefont {Hilbert}(1901)}]{Hilbert1901}%
  \BibitemOpen
  \bibfield  {author} {\bibinfo {author} {\bibfnamefont {D.}~\bibnamefont
  {Hilbert}},\ }\bibfield  {title} {\bibinfo {title} {{\"U}ber {Fl\"a}chen von
  constanter {Gaussscher} {Kr\"u}mmung},\ }\href@noop {} {\bibfield  {journal}
  {\bibinfo  {journal} {Trans. Amer. Math. Soc.}\ }\textbf {\bibinfo {volume}
  {2}},\ \bibinfo {pages} {87} (\bibinfo {year} {1901})}\BibitemShut {NoStop}%
\bibitem [{\citenamefont {Iorio}\ and\ \citenamefont
  {Lambiase}(2014)}]{IorioLambiase2014}%
  \BibitemOpen
  \bibfield  {author} {\bibinfo {author} {\bibfnamefont {A.}~\bibnamefont
  {Iorio}}\ and\ \bibinfo {author} {\bibfnamefont {G.}~\bibnamefont
  {Lambiase}},\ }\bibfield  {title} {\bibinfo {title} {Quantum field theory in
  curved graphene spacetimes, {Lobachevsky} geometry, {Weyl} symmetry,
  {Hawking} effect, and all that},\ }\href@noop {} {\bibfield  {journal}
  {\bibinfo  {journal} {Phys. Rev. D}\ }\textbf {\bibinfo {volume} {90}},\
  \bibinfo {pages} {025006} (\bibinfo {year} {2014})}\BibitemShut {NoStop}%
\bibitem [{\citenamefont {Taioli}\ \emph {et~al.}(2016)\citenamefont {Taioli},
  \citenamefont {Gabbrielli}, \citenamefont {Simonucci}, \citenamefont
  {Pugno},\ and\ \citenamefont {Iorio}}]{Taioli2016}%
  \BibitemOpen
  \bibfield  {author} {\bibinfo {author} {\bibfnamefont {S.}~\bibnamefont
  {Taioli}}, \bibinfo {author} {\bibfnamefont {R.}~\bibnamefont {Gabbrielli}},
  \bibinfo {author} {\bibfnamefont {S.}~\bibnamefont {Simonucci}}, \bibinfo
  {author} {\bibfnamefont {N.~M.}\ \bibnamefont {Pugno}},\ and\ \bibinfo
  {author} {\bibfnamefont {A.}~\bibnamefont {Iorio}},\ }\bibfield  {title}
  {\bibinfo {title} {{Lobachevsky} crystallography made real through carbon
  pseudospheres},\ }\href@noop {} {\bibfield  {journal} {\bibinfo  {journal}
  {J. Phys.: Condens. Matter}\ }\textbf {\bibinfo {volume} {28}},\ \bibinfo
  {pages} {13LT01} (\bibinfo {year} {2016})}\BibitemShut {NoStop}%
\bibitem [{\citenamefont {Morresi}\ \emph {et~al.}(2020)\citenamefont
  {Morresi}, \citenamefont {Binosi}, \citenamefont {Simonucci}, \citenamefont
  {Piergallini}, \citenamefont {Roche}, \citenamefont {Pugno},\ and\
  \citenamefont {Taioli}}]{Morresi2020}%
  \BibitemOpen
  \bibfield  {author} {\bibinfo {author} {\bibfnamefont {T.}~\bibnamefont
  {Morresi}}, \bibinfo {author} {\bibfnamefont {D.}~\bibnamefont {Binosi}},
  \bibinfo {author} {\bibfnamefont {S.}~\bibnamefont {Simonucci}}, \bibinfo
  {author} {\bibfnamefont {R.}~\bibnamefont {Piergallini}}, \bibinfo {author}
  {\bibfnamefont {S.}~\bibnamefont {Roche}}, \bibinfo {author} {\bibfnamefont
  {N.~M.}\ \bibnamefont {Pugno}},\ and\ \bibinfo {author} {\bibfnamefont
  {S.}~\bibnamefont {Taioli}},\ }\bibfield  {title} {\bibinfo {title}
  {Exploring event horizons and {Hawking} radiation through deformed graphene
  membranes},\ }\href@noop {} {\bibfield  {journal} {\bibinfo  {journal} {2D
  Mater.}\ }\textbf {\bibinfo {volume} {7}},\ \bibinfo {pages} {041006}
  (\bibinfo {year} {2020})}\BibitemShut {NoStop}%
\bibitem [{\citenamefont {Girin}\ \emph {et~al.}(2025)\citenamefont {Girin},
  \citenamefont {Bittner}, \citenamefont {Checoury}, \citenamefont {Decanini},
  \citenamefont {Dietz}, \citenamefont {Grigis}, \citenamefont {Lafargue},
  \citenamefont {Zyss}, \citenamefont {Xu}, \citenamefont {Sebbah},\ and\
  \citenamefont {Lebental}}]{Girin2025}%
  \BibitemOpen
  \bibfield  {author} {\bibinfo {author} {\bibfnamefont {H.}~\bibnamefont
  {Girin}}, \bibinfo {author} {\bibfnamefont {S.}~\bibnamefont {Bittner}},
  \bibinfo {author} {\bibfnamefont {X.}~\bibnamefont {Checoury}}, \bibinfo
  {author} {\bibfnamefont {D.}~\bibnamefont {Decanini}}, \bibinfo {author}
  {\bibfnamefont {B.}~\bibnamefont {Dietz}}, \bibinfo {author} {\bibfnamefont
  {A.}~\bibnamefont {Grigis}}, \bibinfo {author} {\bibfnamefont
  {C.}~\bibnamefont {Lafargue}}, \bibinfo {author} {\bibfnamefont
  {J.}~\bibnamefont {Zyss}}, \bibinfo {author} {\bibfnamefont {C.}~\bibnamefont
  {Xu}}, \bibinfo {author} {\bibfnamefont {P.}~\bibnamefont {Sebbah}},\ and\
  \bibinfo {author} {\bibfnamefont {M.}~\bibnamefont {Lebental}},\ }\bibfield
  {title} {\bibinfo {title} {Exploring non-{Euclidean} photonics: Pseudosphere
  microlaser},\ }\href@noop {} {\bibfield  {journal} {\bibinfo  {journal}
  {Phys. Rev. A}\ }\textbf {\bibinfo {volume} {111}},\ \bibinfo {pages}
  {043512} (\bibinfo {year} {2025})}\BibitemShut {NoStop}%
\bibitem [{\citenamefont {Andr{\'e}ief}(1883)}]{Andreief1883}%
  \BibitemOpen
  \bibfield  {author} {\bibinfo {author} {\bibfnamefont {C.}~\bibnamefont
  {Andr{\'e}ief}},\ }\bibfield  {title} {\bibinfo {title} {Note sur une
  relation entre les int{\'e}grales d{\'e}finies des produits des fonctions},\
  }\href@noop {} {\bibfield  {journal} {\bibinfo  {journal} {M{\'e}m. Soc. Sci.
  Phys. Nat. Bordeaux (3)}\ }\textbf {\bibinfo {volume} {2}},\ \bibinfo {pages}
  {1} (\bibinfo {year} {1883})}\BibitemShut {NoStop}%
\bibitem [{\citenamefont {{\v{S}}amaj}\ and\ \citenamefont
  {Percus}(1995)}]{SamajPercus1995}%
  \BibitemOpen
  \bibfield  {author} {\bibinfo {author} {\bibfnamefont {L.}~\bibnamefont
  {{\v{S}}amaj}}\ and\ \bibinfo {author} {\bibfnamefont {J.~K.}\ \bibnamefont
  {Percus}},\ }\bibfield  {title} {\bibinfo {title} {A functional relation
  among the pair correlations of the two-dimensional one-component plasma},\
  }\href@noop {} {\bibfield  {journal} {\bibinfo  {journal} {J. Stat. Phys.}\
  }\textbf {\bibinfo {volume} {80}},\ \bibinfo {pages} {811} (\bibinfo {year}
  {1995})}\BibitemShut {NoStop}%
\bibitem [{\citenamefont {{\v{S}}amaj}(2004)}]{Samaj2004}%
  \BibitemOpen
  \bibfield  {author} {\bibinfo {author} {\bibfnamefont {L.}~\bibnamefont
  {{\v{S}}amaj}},\ }\bibfield  {title} {\bibinfo {title} {Is the
  two-dimensional one-component plasma exactly solvable?},\ }\href@noop {}
  {\bibfield  {journal} {\bibinfo  {journal} {J. Stat. Phys.}\ }\textbf
  {\bibinfo {volume} {117}},\ \bibinfo {pages} {131} (\bibinfo {year}
  {2004})}\BibitemShut {NoStop}%
\bibitem [{\citenamefont {Jancovici}(1982)}]{Jancovici1982}%
  \BibitemOpen
  \bibfield  {author} {\bibinfo {author} {\bibfnamefont {B.}~\bibnamefont
  {Jancovici}},\ }\bibfield  {title} {\bibinfo {title} {Classical {Coulomb}
  systems near a plane wall. {I}},\ }\href@noop {} {\bibfield  {journal}
  {\bibinfo  {journal} {J. Stat. Phys.}\ }\textbf {\bibinfo {volume} {28}},\
  \bibinfo {pages} {43} (\bibinfo {year} {1982})}\BibitemShut {NoStop}%
\bibitem [{\citenamefont {Temme}(1975)}]{Temme1975}%
  \BibitemOpen
  \bibfield  {author} {\bibinfo {author} {\bibfnamefont {N.~M.}\ \bibnamefont
  {Temme}},\ }\bibfield  {title} {\bibinfo {title} {Uniform asymptotic
  expansions of the incomplete gamma functions and the incomplete beta
  function},\ }\href@noop {} {\bibfield  {journal} {\bibinfo  {journal} {Math.
  Comp.}\ }\textbf {\bibinfo {volume} {29}},\ \bibinfo {pages} {1109} (\bibinfo
  {year} {1975})}\BibitemShut {NoStop}%
\bibitem [{\citenamefont {Olver}\ \emph {et~al.}(2010)\citenamefont {Olver},
  \citenamefont {Lozier}, \citenamefont {Boisvert},\ and\ \citenamefont
  {Clark}}]{DLMF}%
  \BibitemOpen
  \bibinfo {editor} {\bibfnamefont {F.~W.~J.}\ \bibnamefont {Olver}}, \bibinfo
  {editor} {\bibfnamefont {D.~W.}\ \bibnamefont {Lozier}}, \bibinfo {editor}
  {\bibfnamefont {R.~F.}\ \bibnamefont {Boisvert}},\ and\ \bibinfo {editor}
  {\bibfnamefont {C.~W.}\ \bibnamefont {Clark}},\ eds.,\ \href@noop {} {\emph
  {\bibinfo {title} {{NIST} Handbook of Mathematical Functions}}}\ (\bibinfo
  {publisher} {Cambridge University Press},\ \bibinfo {address} {Cambridge},\
  \bibinfo {year} {2010})\ \bibinfo {note} {sec.~8.12}\BibitemShut {NoStop}%
\bibitem [{\citenamefont {Shen}\ and\ \citenamefont
  {Forrester}(2026)}]{ShenForrester2026}%
  \BibitemOpen
  \bibfield  {author} {\bibinfo {author} {\bibfnamefont {B.-J.}\ \bibnamefont
  {Shen}}\ and\ \bibinfo {author} {\bibfnamefont {P.~J.}\ \bibnamefont
  {Forrester}},\ }\href@noop {} {\bibinfo {title} {Large {$N$} expansions of
  the partition function for {Coulomb} systems on the surface of a cylinder}}
  (\bibinfo {year} {2026}),\ \Eprint {https://arxiv.org/abs/2609.12487}
  {arXiv:2609.12487} \BibitemShut {NoStop}%
\bibitem [{\citenamefont {Choquard}\ \emph {et~al.}(1980)\citenamefont
  {Choquard}, \citenamefont {Favre},\ and\ \citenamefont {Gruber}}]{CFG1980}%
  \BibitemOpen
  \bibfield  {author} {\bibinfo {author} {\bibfnamefont {P.}~\bibnamefont
  {Choquard}}, \bibinfo {author} {\bibfnamefont {P.}~\bibnamefont {Favre}},\
  and\ \bibinfo {author} {\bibfnamefont {C.}~\bibnamefont {Gruber}},\
  }\bibfield  {title} {\bibinfo {title} {On the equation of state of classical
  one-component systems with long-range forces},\ }\href@noop {} {\bibfield
  {journal} {\bibinfo  {journal} {J. Stat. Phys.}\ }\textbf {\bibinfo {volume}
  {23}},\ \bibinfo {pages} {405} (\bibinfo {year} {1980})}\BibitemShut
  {NoStop}%
\bibitem [{\citenamefont {Metropolis}\ \emph {et~al.}(1953)\citenamefont
  {Metropolis}, \citenamefont {Rosenbluth}, \citenamefont {Rosenbluth},
  \citenamefont {Teller},\ and\ \citenamefont {Teller}}]{Metropolis1953}%
  \BibitemOpen
  \bibfield  {author} {\bibinfo {author} {\bibfnamefont {N.}~\bibnamefont
  {Metropolis}}, \bibinfo {author} {\bibfnamefont {A.~W.}\ \bibnamefont
  {Rosenbluth}}, \bibinfo {author} {\bibfnamefont {M.~N.}\ \bibnamefont
  {Rosenbluth}}, \bibinfo {author} {\bibfnamefont {A.~H.}\ \bibnamefont
  {Teller}},\ and\ \bibinfo {author} {\bibfnamefont {E.}~\bibnamefont
  {Teller}},\ }\bibfield  {title} {\bibinfo {title} {Equation of state
  calculations by fast computing machines},\ }\href@noop {} {\bibfield
  {journal} {\bibinfo  {journal} {J. Chem. Phys.}\ }\textbf {\bibinfo {volume}
  {21}},\ \bibinfo {pages} {1087} (\bibinfo {year} {1953})}\BibitemShut
  {NoStop}%
\end{thebibliography}%

\end{document}